\documentclass[aip,apl,amssymb,reprint]{revtex4-2}%
\usepackage{amsmath}
\usepackage{graphicx}
\usepackage{dcolumn}
\usepackage{bm}
\usepackage[utf8]{inputenc}
\usepackage[T1]{fontenc}
\usepackage{mathptmx}
\usepackage{etoolbox}
\UseRawInputEncoding
\usepackage[misc]{ifsym}

\usepackage{url}
\usepackage[colorlinks]{hyperref}
\hypersetup{%
	plainpages=true,
	breaklinks=true,
	hypertexnames=false,
	pageanchor=true,
	colorlinks=true,
	linkcolor={cyan},
	citecolor={cyan},
	urlcolor={cyan},
	anchorcolor={cyan}
}

\makeatletter
\def\@email#1#2{%
	\endgroup
	\patchcmd{\titleblock@produce}
	{\frontmatter@RRAPformat}
	{\frontmatter@RRAPformat{\produce@RRAP{*#1\href{mailto:#2}{#2}}}\frontmatter@RRAPformat}
	{}{}
}%
\makeatother

\begin{document}
	
\preprint{AIP/123-QED}

\title{Resource-efficient parallel entanglement generation for multinode quantum networks via time-bin multiplexing}
\author{Wenbo Zhang}
\thanks{These authors contributed equally to this work.}
\affiliation{MIIT Key Laboratory of Semiconductor Microstructure and Quantum sensing, School of Physics,  Nanjing University of Science and Technology, Nanjing {\rm 210094}, China}
\author{Jing Zheng}
\thanks{These authors contributed equally to this work.}
\affiliation{MIIT Key Laboratory of Semiconductor Microstructure and Quantum sensing, School of Physics,  Nanjing University of Science and Technology, Nanjing {\rm 210094}, China}
\author{Yimin Wang$^{(\textrm{\Letter})}$}
\affiliation{Communications Engineering College, Army Engineering University, Nanjing {\rm 210007}, China}
\author{Tao Li$^{(\textrm{\Letter})}$}
\affiliation{MIIT Key Laboratory of Semiconductor Microstructure and Quantum sensing, School of Physics,  Nanjing University of Science and Technology, Nanjing {\rm 210094}, China}
\affiliation{Engineering Research Center of Semiconductor Device Optoelectronic Hybrid Integration in Jiangsu Province, Nanjing {\rm 210094}, China}
\email{vivhappyrom@163.com and tao.li@njust.edu.cn }

\date{\today}

\begin{abstract}
Nonlocal entanglement generation among multiple remote quantum nodes provides a critical foundation for a variety of counterintuitive quantum applications. The exponential loss of photons transmitting over optical fibers sets an upper limit for entangling these quantum nodes. Here, we propose a resource-efficient and parallel protocol for entangling multiple remote quantum nodes via time-bin multiplexing. The transmission of a single photon with qudit-encoding in the time-bin mode enables entangling multiple stationary qubits in  parallel, when single photons and individual stationary qubits interfaces are used and photon-state modulations are properly introduced before subsequently impinging the photon into each interface. Our protocol can generate parallel multipartite entanglement among ($N\geq3$) quantum nodes with the dimension of the photonic time bins independent of $N$, exponentially reducing the requirements for the coherence time of the stationary qubits and for the complexity of the photonic modulations. These distinct features make our protocol particularly advantageous for the development of multinode quantum networks.
\end{abstract}

\maketitle

Quantum entanglement exhibits strong nonclassical correlations for various practical applications, ranging from quantum communication and computing to quantum metrology~\cite{xu2020secure,SHENG2022One-step,yan2022factoring,Li2024Heralded,giovannetti2011advances}. Nonlocal entanglement between spatially separated stationary qubits is crucial for implementing distributed quantum computing~\cite{Cirac99Distributed,Jiang2007Distributed,qin2017heralded,Su2024Heralded} and scalable quantum networks~\cite{Ruf2021Quantum,wehner2018quantum,sheng2013hybrid}. Several protocols have been proposed to generate nonlocal entanglement for various physical platforms~\cite{qin2024quantum,xiang2013Hybrid,lodahl2015interfacing,reiserer2015cavity,zhou2018preparing,Qiao2022Generation,xia2016generating, macri2018simple}, and they can generally be classified into three main branches~\cite{northup2014quantum,Munro2015Inside,borregaard2019quantum}: (1)  Generating hybrid entanglement between single photons and stationary qubits and measuring these photons after the interference of them to project stationary qubits into an entangled state~\cite{cabrillo1999creation,yu2007robust,li2016rejecting, Nemoto2014Photonic, hurst2019generating, Pompili2021Realization}; (2) Taking a single photon as a data bus to sequentially entangle each stationary qubit and projecting them into an entangled state with proper measurements of the photon~\cite{duan2004scalable,Hu2008Giant,WangC2011EP, li2018gate,Du2025Heralded}; (3) Mapping photonic entanglement into stationary qubit entanglement~\cite{Liu2021Heralded,tiurev2021fidelity,
Lago-Rivera2021Telecom-heralded,Jones2016Design,Liu2025Deterministic}. 
In all these protocols, the efficiency of generating entanglement between spatially separated stationary qubits is fundamentally limited by photon transmission losses in optical channels. Specifically, the upper bounded efficiency for generating an entangled state decreases exponentially with the transmission distance, proportional to exp$(-\alpha L)$~\cite{northup2014quantum,Munro2015Inside,borregaard2019quantum}, where $\alpha$ is the channel loss rate and $L$ is the channel length.

Recent advances have introduced quantum multiplexing to improve entanglement generation efficiency~\cite{Piparo2019multiplexing,xie2021quantum,Wang2025Heralded}. A notable protocol was proposed to generate two pairs of stationary Bell states by encoding a single photon in both the polarization and time-bin qubits~\cite{Piparo2019multiplexing}. A general quantum multiplexing protocol for simultaneously generating multiple~($M>2$) pairs of stationary Bell states was presented by using a time-bin qudit~\cite{Erhard2020high-dimensional} and a polarization qubit of a single photon~\cite{xie2021quantum}. 
These protocols offer a significant advantage in terms of parallel entanglement generation~\cite{McIntyre2025Loss-tolerant}, with the efficiency for generating one pair of entanglement scaling as exp$(-\alpha L/M)$, thereby improving the overall efficiency. 
Additionally, a compact and equally efficient Bell entangling protocol has been proposed~\cite{zheng2022entanglement}, using a single photon encoded in a time-bin qudit~\cite{liu2024error}, with uniform photonic time-bin modulations across two nodes. More recently, a multiplexing protocol for generating stationary multipartite Greenberger-Horne-Zeilinger (GHZ) states was introduced~\cite{zhou2023parallel}, capable of generating $M$ $N$-qubit GHZ states in parallel. However, the dimension of the photonic Hilbert space scales exponentially as $2^{(M-1)N}$, which grows rapidly with $N$.

Here we propose a compact and resource-efficient protocol for generating multiple  multipartite stationary GHZ states in parallel. Our protocol utilizes a single photon encoded in a qudit of dimension  $d=2^{M}$, which enables the generation of  $M$ stationary GHZ states among $N$ spatially separated quantum nodes. The dimension of the photon space depends only on $M$ and remains independent of the number of quantum nodes, significantly reducing the coherence time requirements of stationary qubits. This is particularly advantageous for constructing large-scale quantum networks involving a large number of quantum nodes. Moreover, our protocol involves $M$ distinct time-bin modulations, determined by the relative positions of the qubits within each node, in contrast to the $(NM - 1)$ distinct and technically demanding modulations required in previous protocols~\cite{zhou2023parallel}. Therefore, our parallel protocol is compact and resource-efficient for implementing scalable multinode quantum networks and distributed quantum computing.

\begin{figure}
    \centering
    \includegraphics[width=0.98\linewidth]{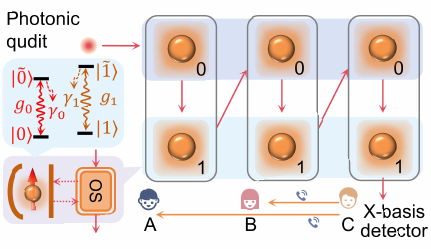}
    \caption{
     Schematic of the compact protocol for generating two \emph{three}-qubit GHZ states in parallel. Each node has two cavity-qubit units, each controlled by an optical switch to control the photon-qubit interaction. The photonic time-bin qudit interacts probabilistically with the qubits in node A, and then travels to nodes B and C, where it interacts with the qubits in the same manner. At the final node, the photon is measured by an $X$-basis detector, and upon successful detection, the measurement outcome is communicated to the other nodes, thereby enabling the generation of two \emph{three}-qubit GHZ states across three nodes.
}
    \label{fig1}
\end{figure}

Consider $N$ nodes in a quantum network and each node has $M$ stationary qubits, each coupling to a single-sided cavity, that can efficiently interact with single photons. 
A single photon encodes a time-bin qudit with  dimension $d$, corresponding to the Hilbert space of $M$ time-bin qubits. By properly tuning the interaction between the photon and stationary qubits, each time-bin qubit can sequentially entangle $N$ stationary qubits. As a result, $M$ nonlocal $N$-qubit GHZ states can be generated simultaneously in a heralded way, up to local single-qubit transformations. This entanglement generation is signaled by the successful measurement of the single photon in the   $X$-basis, as shown in Fig.~\ref{fig1}.

We assume that all stationary qubits are in the state $|0\rangle$ and a $d$-dimensional photon is in the superposition state~\cite{aharonovich2016solid}
\begin{align}
    \left | \phi  \right \rangle =\frac{1}{\sqrt{d}}  \sum_{l=0}^{d-1}\left | l \right \rangle, \label{ph state}
\end{align}
where $|l\rangle$ denotes the $l{\rm th}$ time bin.  In practice, the decimal $l$ can be converted into an $M$-bit binary series, which determines the qubits that interact with the photon, inducing a bit flip on the corresponding qubits. Specifically, the binary representation $(l)_D=(i_{M-1}..i_1i_0)_B$ with $i_m\in \left \{ 0,1 \right\}$ indicates the interaction pattern, and the subscript $m\in\left \{0,1..M-1  \right \}$ denotes the $m$th qubit at each node.

Upon reaching each node, the photon is impinged into the $m$th cavity-qubit unit via an optical switch~(OS) placed before the cavity. The OS can be realized using fast electro-optic modulators, such as Mach-Zehnder interferometers~\cite{Borregaard2020One-Way}, which support sub-nanosecond operation with low insertion loss~\cite{Wang2018Integrated} and are well suited for controlling time-bin photonic qudits. If the corresponding binary bit $i_m=1$, the photon interacts with the $m$th stationary qubit, flipping its  state (i.e., $\left | 0_m  \right \rangle\Rightarrow\left | 1_m\right\rangle$);  Otherwise, the photon is transmitted directly by the OS to the next cavity-qubit unit. After the photon interacts sequentially with all stationary qubits, the system, consisting of the photon and   $NM$ qubits, evolves into a hybrid entangled state
\begin{align}
	\left | \Psi  \right \rangle & = \frac{1}{\sqrt{d} }\sum_{l = 0}^{d-1} \left | l^{\tilde{N}}_D  \right \rangle,
\end{align}
where $\left | l^{\tilde{N}}_D\right \rangle=\left |l \right\rangle_{}\left | i_{M-1}..i_1i_0  \right \rangle^{\otimes N}$ denotes the state in which the $m$th qubit across all $N$ nodes shares the same state $|i_m\rangle$, exclusively determined by the photon time-bin state $|l\rangle$.

The photon is then measured by a generalized $X$-basis detector to convert the hybrid entanglement between the time-bin qudit and the $NM$ stationary qubits into multipartite entanglement among the qubits at identical positions across the $N$ nodes. Specifically, 
a generalized $X$-basis measurement~\cite{zheng2022entanglement} on the time-bin qudit collapses the photon's state into an $X$-basis state via the Fourier transform
\begin{align}
    \left | l  \right \rangle_{}=\frac{1}{\sqrt{d}}\sum_{k=0}^{d-1} \theta^{kl} \left | k  \right \rangle_{},\label{fourtrans}
\end{align}
where $\theta = \exp(2\pi i/d)$, and $\left | k \right \rangle$ denotes the $k$th $X$-basis state. Upon measuring the photon in the state $\left | k \right \rangle$, the $NM$ stationary qubits are projected into the state
\begin{align}
	\left | \Psi_k   \right \rangle  =&\frac{1}{\sqrt{d}}  \sum_{l  = 0}^{d-1}\theta^{-kl}\left | i_{M-1}..i_1i_0  \right \rangle^{\otimes N} \nonumber  \\ 
	  =& \frac{1}{\sqrt{d}} \bigotimes_{\tilde{m}=1}^M (| 00..0\rangle +\theta^{-2^{\tilde{m}-1}k} | 11..1\rangle)_{\tilde{m}},
\end{align}
which heralds the simultaneous generation of $M$ GHZ states among $N$ remote nodes. These $M$ GHZ states can be converted into the  GHZ state $|\Psi_s\rangle= \frac{1}{\sqrt{2} }\left (\left | 00..0   \right \rangle +\left | 11..1 \right \rangle \right )$ by applying a local phase operation $\varphi=\theta ^{2^{\tilde{m}-1}k}$ to the $\tilde{m}$th qubit at node $N$. No additional communication is required among the remaining $(N-1)$ nodes.

For a specific example with $M  = 2$ and  $N  = 3$, our protocol can be detailed as follows. All stationary qubits are initially prepared in the state $\left|0\right\rangle$, and the photonic time-bin qudit is initialized in the state $\left |\tilde{\phi}\right \rangle =\frac{1}{2}\sum_{l=0}^{3}\left | l  \right \rangle$. The four decimal time-bin indices $l \in \{0, 1, 2, 3\}$ are mapped directly to two-bit binary strings $(\alpha\beta)_{B}$. Here, the binary values $\alpha, \beta \in \{0, 1\}$ dictate the OS, directing the photon to interact with the first ($\alpha=1$) or second ($\beta=1$) cavity-qubit unit at each node.

After interacting with two qubits  $A_0$ and $A_1$ in node A, the state of the photon and  $A_0A_1$  evolves into $\left | \Psi _A \right \rangle=\frac{1}{2}\sum_{l=0}^{3}\left |l \right\rangle_{}\left |i_1i_0  \right \rangle$. The photon is then transmitted to nodes B and C, and  interacts with stationary qubits there in the same manner as in node A. The hybrid entangled state between the photon and six stationary qubits evolves to 
\begin{align}
\left|\tilde{\Psi}\right\rangle
= &\frac{1}{2}(|0\rangle|00\rangle^{\otimes 3}
+|1\rangle|01\rangle^{\otimes 3}
+|2\rangle|10\rangle^{\otimes 3} \notag\\
&+|3\rangle|11\rangle^{\otimes 3}).
\end{align}

Upon measurement of the photon in the state $\left|k\right \rangle$, the six stationary qubits are projected onto the state
\begin{align}
	\left | \tilde{\Psi}_k\right \rangle=& \frac{1}{2}( \left | 0_{A_0}0_{B_0}0_{C_0} \right \rangle + \theta^{-k} \left | 1_{A_0}1_{B_0}1_{C_0} \right \rangle )_{\tilde{1}}  \notag\\
	& \otimes ( \left | 0_{A_1}0_{B_1}0_{C_1}  \right \rangle + \theta^{-2k} \left | 1_{A_1}1_{B_1}1_{C_1}  \right \rangle )_{\tilde{2}},   
\end{align} 
where $\theta = \exp(\pi i/2)$, and $\left | \tilde{\Psi}_k\right \rangle$ denotes two GHZ states among the qubits ${A_0}{B_0}{C_0}$ and ${A_1}{B_1}{C_1}$ simultaneously.

The photon-induced state flip of a stationary qubit can be implemented by a controlled-phase flip~(CPF) operation in combination with two Hadamard operations applied before and after the CPF~\cite{Ruf2021Quantum,Knaut2024Entanglement}. The CPF operation can be implemented using a variety of photon-matter interfaces, such as color centers coupled to optical cavities, neutral atoms in microcavities, or trapped-ion-cavity systems~\cite{xiang2013Hybrid,lodahl2015interfacing,reiserer2015cavity,borregaard2019quantum}. Here we focus on the implementation based on color centers coupled to optical cavities~\cite{Ruf2021Quantum}.
The CPF operation involves the scattering of a photon by a single-sided cavity coupled to a four-level atom, as shown in Fig.~\ref{fig1}. The atom has two ground states $\left | 0  \right \rangle$ and  $\left | 1  \right \rangle$, and two excited states $\left | \tilde{0}  \right \rangle$ and  $\left | \tilde{1}  \right \rangle$, with decay rates $\gamma_0$ and $\gamma_1$, respectively. The dipole-allowed transition  $\left | s  \right \rangle\leftrightarrow\left | \tilde{s}  \right \rangle$ for $s=0,1$ interacts with the cavity mode at a real coupling rate $g_s$, and  is  detuned from the cavity resonant mode by $\Delta_s$. The Hamiltonian of the system is described by
\begin{align}
	\hat{H} & = \sum_{s = 0}^{1}[\Delta _{s} \left | \tilde{s}  \right \rangle  \left  \langle \tilde{s} \right |+ (g_s\left | \tilde{s}  \right \rangle\left \langle s \right |\hat{c} + \textrm{H.C.})],     
\end{align}
where $\hat{c}$ is the cavity mode that couples to the collected~(leakage) mode at a rate $\kappa_a$~($\kappa_1$). The dynamic equation for the cavity mode $\hat{c}$ is 
\begin{align}
\dot{\hat{c} }  = i[\hat{H},\hat{c}]-\frac{\kappa\hat{c}}{2}+ \sqrt{\kappa _a}\hat{a}_{\rm in},
\end{align}
where $\kappa=\kappa_a+\kappa_1$, and $\hat{a}_{\rm in}$ is the input mode related to the output mode via the input-output relation $\hat{a}_{\rm out}  = \hat{a}_{\rm in}-\sqrt{\kappa _a} \hat{c}$. 

The  state-dependent reflection coefficient $r_s$~($s=0,~1$) for a single photon with frequency detuning $\omega$ from the cavity resonant mode can be described as~\cite{reiserer2015cavity} 
\begin{eqnarray}    
r_s =1-\frac{2\kappa_a/\kappa(i\Delta_s+1)}{(i\Delta_s +1)(i\Delta_c+1)+C_s},
	\label{rcoe}
\end{eqnarray}
where $C_s=4g_s^{2 } /\kappa\gamma_s$ is the cooperativity,  $\Delta_{s}=2(\omega_{s}-\omega)/\gamma_s$,
and  $\Delta_c=-2\omega/\kappa$ denote the effective detunings. 
For $\kappa_a/\kappa=0.98$, $C_0=C_1=45$, $\Delta_c=0.3$, $\Delta_0=0$, and
$\Delta_1=150$, one obtains $r_0\simeq -r_1\simeq0.96$ with a phase difference of $\pi$~\cite{xie2021quantum}.
This yields unit fidelity but reduces the efficiency of each CPF operation to $\eta_0=r_0^2$, resulting in an overall reduction factor $\eta_1=\eta_0^{NM/2}$. 

Besides nonideal scattering, decoherence of the stationary qubits further degrades the entanglement generation performance. The evolution of each stationary qubit is modeled by
\begin{equation}\label{key}
\hat{\rho}(t)=\sum_{i=\pm}\sum_{j=0}^{3} [\hat{A}_i(t)\hat{B}_j(t)]\hat{\rho}[\hat{A}_i(t)\hat{B}_j(t)]^{\dagger}.
\end{equation}
The Kraus operators describing damping are $\hat{B}_{0}(t)=\frac{1}{\sqrt2}(|0\rangle\langle0|+\sqrt{\mu_1}|1\rangle\langle1|)$, $\hat{B}_{1}(t)=\frac{1}{\sqrt2}(\sqrt{\mu_1}|0\rangle\langle0|+|1\rangle\langle1|)$,
and $\hat{B}_2(t)=\hat{B}_3^\dagger(t)=\sqrt{\frac{1-\mu_1}{2}}|0\rangle\langle1|$ 
with $\mu_1=\textrm{exp}(-t/T_1)$. The dephasing  process is characterized by 
$\hat{A}_{\pm}(t)=\sqrt{\frac{1\pm\mu_2}{2}}(|0\rangle\langle0|\pm|1\rangle\langle1|)$
with $\mu_2=\textrm{exp}(-t/T_2)$. Here $T_1$ and $T_2$ denote the relaxation and dephasing times, respectively. We assume that the stationary qubit at each node is first transformed to $|+\rangle=(|0\rangle+|1\rangle)/\sqrt2$ via a Hadamard operation before the photon arrives, and converted back to the $Z$ basis after photon detection.

To estimate the protocol duration, we assume that the distance between adjacent nodes is $L_0$.
The photon transmission time is therefore $t_0=L_0/c$, where $c=2\times10^8~\mathrm{m/s}$.
Assuming all nodes are arranged linearly, the classical communication time is $(N-1)t_0$.
The total time required to generate $M$ $N$-qubit GHZ states is thus $2(N-1)t_0$.
The time for local operations is neglected for simplicity, since it is much shorter than $t_0$ for $L_0>3$ km and a time-bin separation of $\sim0.1~\mu\mathrm{s}$~\cite{Knaut2024Entanglement}. 

The $X$-basis measurement for the photonic time-bin qudit can be implemented using an interferometer. Its size and phase noise scale  with the number of GHZ states $M$. The resulting overall dephasing effect can be modeled as $\Lambda_D(\rho) = \lambda_d\rho + (1 - \lambda_d) \sum_{k=0}^{d-1} \hat{\sigma}_{kk} \rho \hat{\sigma}_{kk}$, where $\hat{\sigma}_{kk} = |k\rangle\langle k|$ and $\lambda_d = \exp(-x^2 / 2)$ with $x = 0.1M$. This corresponds to an attenuation of the off-diagonal density-matrix elements by a factor $\lambda_d$~\cite{zheng2022entanglement}. For the interferometer containing $n=(d-1)d/2$ fiber loops, each introducing a loss $\eta_l$, the average efficiency of the interferometer is $\eta_2=(1-\eta_l)^{(d-1)/2}$. 

A practical OS transforms an input photon as $\lvert \mathrm{in} \rangle \rightarrow \sqrt{1-e_{\mathrm{os}}}\,\lvert \mathrm{out} \rangle_{1} + \sqrt{e_{\mathrm{os}}}\,\lvert \mathrm{out} \rangle_{2} $ with efficiency $\eta_{\rm os}$ and error rate $e_{\mathrm{os}}$. Here $|\mathrm{out}\rangle_{1,2}$ denote the desired and erroneous outputs. Its effect on the fidelity of the $X$-basis measurement can be passively suppressed by postselecting the appropriate detection time of the single-photon detectors~\cite{zheng2022entanglement}.
This leads to an overall efficiency reduction factor of $\eta_3=[\eta_{\rm os}(1-e_{\mathrm{os}})]^{M(N+1)}$. 

The average efficiency for generating $M$ $N$-qubit GHZ states is 
\begin{eqnarray}
\bar{\eta}_{M,N}=\eta_1\eta_2\eta_3\eta_4, 
\end{eqnarray}
where $\eta_4=\exp[-\alpha(N-1)L_0]$ is the channel transmission efficiency with $\alpha=\frac{1}{20}~{\rm km}^{-1}$.
The corresponding average fidelity is~\cite{johansson2013qutip,Zhang2021Efficient}
\begin{eqnarray}
\bar{F}_{M,N}=\frac{1}{d}\sum_{k=0}^{d-1}{\rm tr}(\sqrt{\rho_k}\tilde{\rho}_k\sqrt{\rho_k})^{\frac{1}{M}}, 
\end{eqnarray}
where $\rho_k=|\Psi_k\rangle\langle\Psi_k|$ is the ideal target state corresponding to the photon measurement outcome $|k\rangle$, and $\tilde{\rho}_k$ is the practical density matrix including decoherence, imperfect cavity scattering and OS, interferometric phase errors, and single-photon qudit noise. 

\begin{figure}
	\centering
	\includegraphics[width=1\linewidth]{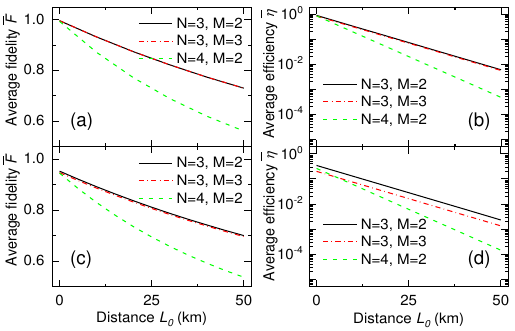}
	\caption{Average fidelities $\bar{F}_{M,N}$ and efficiencies $\bar{\eta}_{M,N}$ as functions of the distance $L_0$ for three representative parameter sets $(M,N)=(2,3),~(3,3),~(2,4)$ with $T_1=2T_2=10~\mathrm{ms}$.
(a) and (b) show the average fidelities and efficiencies for ideal optical switches, photonic qudits, and interferometers, respectively.
(c) and (d) present the corresponding results for practical devices with $\eta_{\rm os}=0.9$, $e_{\rm os}=0.01$, $\eta_l=0.01$, and $x=0.1M$.}
	\label{fig2}
\end{figure}

The average fidelities $\bar{F}_{M,N}$ and efficiencies $\bar{\eta}_{M,N}$ as functions of the distance $L_0$ are shown in Fig.~\ref{fig2}. 
For  ideal OSs, a single-photon source, and an interferometer with $\eta_{\rm os}=1$, $e_{\mathrm{os}}=0$, $\eta_l=0$, and $x=0$, the average fidelities approach $\bar{F}_{M,N}\simeq1$ for $L_0\rightarrow0$, as shown in Fig.~\ref{fig2}(a). 
For generating $M=3$ \emph{three}-qubit GHZ states, we obtain $\bar{F}_{3,3}>0.85$ for $T_1=2T_2=10$ ms and $L_0<25$ km, which is nearly identical to the case of generating $M=2$ \emph{three}-qubit GHZ states. For generating $M=2$ \emph{four}-qubit GHZ states we obtain $\bar{F}_{2,4}>0.73$ for $L_0<25$ km. 
For practical OSs with $\eta_{\rm os}=0.9$ and $e_{\mathrm{os}}=0.01$ and a practical interferometer with $\eta_l=0.01$ and $x=0.1M$, we still have $\bar{F}_{3,3}>0.81$ and $\bar{F}_{2,4}>0.70$ for $L<25$ km for a practical photonic qudit. Here we model the noisy photonic qudit state as
\begin{eqnarray}
|\psi\rangle_{\text{noisy}} = \frac{1}{\sqrt{C_d}} \sum_{j=0}^{d - 1} (1 + \alpha_j) e^{i\theta_j} |j\rangle,
\end{eqnarray}
where the normalization factor is $C_d = \sum_{j=0}^{d - 1} |1 + \alpha_j|^2$, and $\alpha_j \sim {N}(0, \sigma_a^2)$ and $\theta_j \sim {N}(0, \sigma_p^2)$ denote Gaussian-distributed amplitude and phase fluctuations, respectively. The fluctuations are assumed to be independent for different time bins with $\sigma_a = \sigma_p = 0.1$~\cite{zheng2022entanglement}.

In addition to the error sources explicitly included in our model, several practical imperfections may, in principle, affect the fidelity of the protocol. Dark counts in state-of-the-art single-photon detectors are typically below $10^{-8}$ per detection window~\cite{xu2020secure}, and thus their contribution is negligible compared to the intrinsic success probability of the protocol. Multiphoton emission is strongly suppressed for cavity-emitter-based single-photon sources~\cite{lodahl2015interfacing,reiserer2015cavity}, and its effect remains negligible for attenuated coherent states with mean photon number $|\alpha|^2 \lesssim 0.05$~\cite{reiserer2015cavity,Guo2023Heralded}. 

Nonuniform cavity parameters, such as variations in cooperativity and detuning across different nodes and qubits~\cite{borregaard2019quantum}, can impact protocol performance. However, our numerical analyses demonstrate that within realistic parameter ranges (e.g., $C \in [40, 50]$, $\Delta_0 \in [-1, 1]$, and $\Delta_1 \in [140, 160]$), the amplitude and phase of the reflection coefficients remain highly stable, inducing only negligible deviations in the photon-qubit interactions. Even in a worst-case scenario combining reduced cooperativity with maximum detuning offsets, the protocol maintains robust fidelities over relevant transmission distances with  $\overline{F}_{3,3} > 0.80$ and $\overline{F}_{2,4} > 0.69$ for $L_0 < 25$ km.

Chromatic and temporal distinguishability of time-bin modes after fiber propagation, arising from dispersion and temperature-induced path-length fluctuations, can be effectively mitigated using dispersion compensation and active phase stabilization~\cite{Fitzke2022Scalable}. The arrival-time drifts can be compensated via automatic phase calibration and clock recovery, reducing the associated error rate to $\sim 0.4\%$ over distances exceeding $60$~km~\cite{Fitzke2022Scalable}. The temporal broadening remains well below the time-bin separation. Therefore, the residual impact of the above imperfections is expected to be significantly smaller than that of dominant noise sources such as qubit decoherence and measurement-induced dephasing.

The corresponding average efficiencies of our protocol for generating $M$ $N$-qubit GHZ states across $N$ spatially separated nodes are shown in Figs.~\ref{fig2}(b) and \ref{fig2}(d) for ideal and practical setups, respectively. 
In the practical case, we obtain  $\bar{\eta}_{3,3}>1.37\%$ and $\bar{\eta}_{2,4}>0.53\%$ for $L_0<25$ km, which can be improved to $\bar{\eta}_{3,3}>5.68\%$ and $\bar{\eta}_{2,4}>1.70\%$ by using an ideal interferometer and OSs.

The average time required to complete the protocol is approximately
$t=2(N-1) L_0 /(c\,\bar{\eta}_{M,N})$. To generate $M$ $N$-qubit GHZ states with fidelities above the threshold value of $0.5$, which is a sufficient condition for witnessing genuine multipartite entanglement~\cite{Song2019Generation}, the minimum coherence time required for the stationary qubits approaches $t_{\min}\simeq t$~\cite{Borregaard2020One-Way}. 
In contrast, conventional protocols that require the successful transmission of $M$ photons demand a minimum coherence time proportional to
$2(N-1) L_0 /(c\,\eta_4^m)$.
Therefore, our protocol substantially relaxes the coherence-time requirement of stationary qubits, reducing it by several orders of magnitude compared with conventional schemes, even for node separations exceeding $L_0>25$ km and for $M>2$. 
This advantage significantly mitigates the experimental challenges associated with implementing quantum tasks involving multiple spatially separated multi-qubit states.

Beyond relaxing coherence requirements, our protocol represents a conceptual and practical advance over existing multiplexing frameworks~\cite{Piparo2019multiplexing,xie2021quantum,Wang2025Heralded} by defining a new operational regime for time-bin encoding. By using probabilistic photon-qubit interactions, our scheme requires only half as many scattering events as deterministic protocols where a photon must sequentially interact with every stationary qubit~\cite{Piparo2019multiplexing,xie2021quantum,Wang2025Heralded}. This fundamentally eliminates the need for strict global temporal synchronization or dynamically controlled delays between successive photon-cavity-qubit interactions~\cite{zhou2023parallel}. Furthermore, the factorization of the interaction pattern drastically reduces experimental control complexity; the system requires only $M$ distinct time-bin modulations, in contrast to the highly demanding $(NM-1)$ modulations required by the previous parallel GHZ-state generation approach~\cite{zhou2023parallel}.

In summary, we have proposed a compact and resource-efficient protocol capable of generating multiple GHZ states among arbitrarily many spatially separated stationary qubits. By using probabilistic photon-qubit interactions, the dimension of the photonic time-bin qudit remains strictly fixed at $d=2^M$ to generate $M$ $N$-qubit GHZ states in parallel. This feature allows additional quantum nodes to be incorporated into the network without modifying the photonic modulations performed at preceding nodes. This parallel architecture significantly reduces the experimental control overhead and drastically relaxes the coherence-time requirements on stationary qubits. Therefore, our protocol provides a highly scalable and practically efficient route toward the realization of large-scale multinode quantum networks.

We thank Zehui Guo for helpful discussions. This work was supported by the National Natural Science Foundation of
China (Grant No. 11904171). 

\section*{AUTHOR DECLARATIONS}
\subsection*{Conflict of Interest}
The authors have no conflicts to disclose.
\section*{DATA AVAILABILITY}
The data that support the findings of this study are available from the corresponding authors upon reasonable request.

\begin{thebibliography}{56}%
\makeatletter
\providecommand \@ifxundefined [1]{%
 \@ifx{#1\undefined}
}%
\providecommand \@ifnum [1]{%
 \ifnum #1\expandafter \@firstoftwo
 \else \expandafter \@secondoftwo
 \fi
}%
\providecommand \@ifx [1]{%
 \ifx #1\expandafter \@firstoftwo
 \else \expandafter \@secondoftwo
 \fi
}%
\providecommand \natexlab [1]{#1}%
\providecommand \enquote  [1]{``#1''}%
\providecommand \bibnamefont  [1]{#1}%
\providecommand \bibfnamefont [1]{#1}%
\providecommand \citenamefont [1]{#1}%
\providecommand \href@noop [0]{\@secondoftwo}%
\providecommand \href [0]{\begingroup \@sanitize@url \@href}%
\providecommand \@href[1]{\@@startlink{#1}\@@href}%
\providecommand \@@href[1]{\endgroup#1\@@endlink}%
\providecommand \@sanitize@url [0]{\catcode `\\12\catcode `\$12\catcode
  `\&12\catcode `\#12\catcode `\^12\catcode `\_12\catcode `\%12\relax}%
\providecommand \@@startlink[1]{}%
\providecommand \@@endlink[0]{}%
\providecommand \url  [0]{\begingroup\@sanitize@url \@url }%
\providecommand \@url [1]{\endgroup\@href {#1}{\urlprefix }}%
\providecommand \urlprefix  [0]{URL }%
\providecommand \Eprint [0]{\href }%
\providecommand \doibase [0]{https://doi.org/}%
\providecommand \selectlanguage [0]{\@gobble}%
\providecommand \bibinfo  [0]{\@secondoftwo}%
\providecommand \bibfield  [0]{\@secondoftwo}%
\providecommand \translation [1]{[#1]}%
\providecommand \BibitemOpen [0]{}%
\providecommand \bibitemStop [0]{}%
\providecommand \bibitemNoStop [0]{.\EOS\space}%
\providecommand \EOS [0]{\spacefactor3000\relax}%
\providecommand \BibitemShut  [1]{\csname bibitem#1\endcsname}%
\let\auto@bib@innerbib\@empty
\bibitem [{\citenamefont {Xu}\ \emph {et~al.}(2020)\citenamefont {Xu},
  \citenamefont {Ma}, \citenamefont {Zhang}, \citenamefont {Lo},\ and\
  \citenamefont {Pan}}]{xu2020secure}%
  \BibitemOpen
  \bibfield  {author} {\bibinfo {author} {\bibfnamefont {F.}~\bibnamefont
  {Xu}}, \bibinfo {author} {\bibfnamefont {X.}~\bibnamefont {Ma}}, \bibinfo
  {author} {\bibfnamefont {Q.}~\bibnamefont {Zhang}}, \bibinfo {author}
  {\bibfnamefont {H.-K.}\ \bibnamefont {Lo}},\ and\ \bibinfo {author}
  {\bibfnamefont {J.-W.}\ \bibnamefont {Pan}},\ }\bibfield  {title} {\emph
  {\bibinfo {title} {Secure quantum key distribution with realistic devices}},\
  }\href {https://doi.org/10.1103/RevModPhys.92.025002} {\bibfield  {journal}
  {\bibinfo  {journal} {Rev. Mod. Phys.}\ }\textbf {\bibinfo {volume} {92}},\
  \bibinfo {pages} {025002} (\bibinfo {year} {2020})}\BibitemShut {NoStop}%
\bibitem [{\citenamefont {Sheng}\ \emph {et~al.}(2022)\citenamefont {Sheng},
  \citenamefont {Zhou},\ and\ \citenamefont {Long}}]{SHENG2022One-step}%
  \BibitemOpen
  \bibfield  {author} {\bibinfo {author} {\bibfnamefont {Y.-B.}\ \bibnamefont
  {Sheng}}, \bibinfo {author} {\bibfnamefont {L.}~\bibnamefont {Zhou}},\ and\
  \bibinfo {author} {\bibfnamefont {G.-L.}\ \bibnamefont {Long}},\ }\bibfield
  {title} {\emph {\bibinfo {title} {One-step quantum secure direct
  communication}},\ }\href
  {https://doi.org/https://doi.org/10.1016/j.scib.2021.11.002} {\bibfield
  {journal} {\bibinfo  {journal} {Sci. Bull.}\ }\textbf {\bibinfo {volume}
  {67}},\ \bibinfo {pages} {367} (\bibinfo {year} {2022})}\BibitemShut
  {NoStop}%
\bibitem [{\citenamefont {Yan}\ \emph {et~al.}(2022)\citenamefont {Yan},
  \citenamefont {Tan}, \citenamefont {Wei}, \citenamefont {Jiang},
  \citenamefont {Wang}, \citenamefont {Wang}, \citenamefont {Luo},
  \citenamefont {Duan}, \citenamefont {Liu}, \citenamefont {Shi}, \citenamefont
  {Fei}, \citenamefont {Meng}, \citenamefont {Han}, \citenamefont {Shan},
  \citenamefont {Chen}, \citenamefont {Zhu}, \citenamefont {Zhang},
  \citenamefont {Jin}, \citenamefont {Li}, \citenamefont {Song}, \citenamefont
  {Wang}, \citenamefont {Ma}, \citenamefont {Wang},\ and\ \citenamefont
  {Long}}]{yan2022factoring}%
  \BibitemOpen
  \bibfield  {author} {\bibinfo {author} {\bibfnamefont {B.}~\bibnamefont
  {Yan}} \emph {et~al.},\ }\href {https://arxiv.org/abs/2212.12372} {\bibinfo
  {title} {Factoring integers with sublinear resources on a superconducting
  quantum processor}},\ \Eprint
  {https://arxiv.org/abs/2212.12372} {arXiv:2212.12372 [quant-ph]} \BibitemShut
  {NoStop}%
\bibitem [{\citenamefont {Li}\ \emph {et~al.}(2024)\citenamefont {Li},
  \citenamefont {Xie}, \citenamefont {Li}, \citenamefont {Liang}, \citenamefont
  {Li},\ and\ \citenamefont {Li}}]{Li2024Heralded}%
  \BibitemOpen
  \bibfield  {author} {\bibinfo {author} {\bibfnamefont {J.}~\bibnamefont
  {Li}}, \bibinfo {author} {\bibfnamefont {Z.}~\bibnamefont {Xie}}, \bibinfo
  {author} {\bibfnamefont {Y.}~\bibnamefont {Li}}, \bibinfo {author}
  {\bibfnamefont {Y.}~\bibnamefont {Liang}}, \bibinfo {author} {\bibfnamefont
  {Z.}~\bibnamefont {Li}},\ and\ \bibinfo {author} {\bibfnamefont
  {T.}~\bibnamefont {Li}},\ }\bibfield  {title} {\emph {\bibinfo {title}
  {Heralded entanglement between error-protected logical qubits for
  fault-tolerant distributed quantum computing}},\ }\href
  {https://doi.org/10.1007/s11433-023-2245-9} {\bibfield  {journal} {\bibinfo
  {journal} {Sci. China-Phys. Mech. Astron.}\ }\textbf {\bibinfo {volume}
  {67}},\ \bibinfo {pages} {220311} (\bibinfo {year} {2024})}\BibitemShut
  {NoStop}%
\bibitem [{\citenamefont {Giovannetti}\ \emph {et~al.}(2011)\citenamefont
  {Giovannetti}, \citenamefont {Lloyd},\ and\ \citenamefont
  {Maccone}}]{giovannetti2011advances}%
  \BibitemOpen
  \bibfield  {author} {\bibinfo {author} {\bibfnamefont {V.}~\bibnamefont
  {Giovannetti}}, \bibinfo {author} {\bibfnamefont {S.}~\bibnamefont {Lloyd}},\
  and\ \bibinfo {author} {\bibfnamefont {L.}~\bibnamefont {Maccone}},\
  }\bibfield  {title} {\emph {\bibinfo {title} {Advances in quantum
  metrology}},\ }\href {https://doi.org/10.1038/nphoton.2011.35} {\bibfield
  {journal} {\bibinfo  {journal} {Nat. Photonics}\ }\textbf {\bibinfo {volume}
  {5}},\ \bibinfo {pages} {222} (\bibinfo {year} {2011})}\BibitemShut {NoStop}%
\bibitem [{\citenamefont {Cirac}\ \emph {et~al.}(1999)\citenamefont {Cirac},
  \citenamefont {Ekert}, \citenamefont {Huelga},\ and\ \citenamefont
  {Macchiavello}}]{Cirac99Distributed}%
  \BibitemOpen
  \bibfield  {author} {\bibinfo {author} {\bibfnamefont {J.~I.}\ \bibnamefont
  {Cirac}}, \bibinfo {author} {\bibfnamefont {A.~K.}\ \bibnamefont {Ekert}},
  \bibinfo {author} {\bibfnamefont {S.~F.}\ \bibnamefont {Huelga}},\ and\
  \bibinfo {author} {\bibfnamefont {C.}~\bibnamefont {Macchiavello}},\
  }\bibfield  {title} {\emph {\bibinfo {title} {Distributed quantum computation
  over noisy channels}},\ }\href {https://doi.org/10.1103/PhysRevA.59.4249}
  {\bibfield  {journal} {\bibinfo  {journal} {Phys. Rev. A}\ }\textbf {\bibinfo
  {volume} {59}},\ \bibinfo {pages} {4249} (\bibinfo {year}
  {1999})}\BibitemShut {NoStop}%
\bibitem [{\citenamefont {Jiang}\ \emph {et~al.}(2007)\citenamefont {Jiang},
  \citenamefont {Taylor}, \citenamefont {S\o{}rensen},\ and\ \citenamefont
  {Lukin}}]{Jiang2007Distributed}%
  \BibitemOpen
  \bibfield  {author} {\bibinfo {author} {\bibfnamefont {L.}~\bibnamefont
  {Jiang}}, \bibinfo {author} {\bibfnamefont {J.~M.}\ \bibnamefont {Taylor}},
  \bibinfo {author} {\bibfnamefont {A.~S.}\ \bibnamefont {S\o{}rensen}},\ and\
  \bibinfo {author} {\bibfnamefont {M.~D.}\ \bibnamefont {Lukin}},\ }\bibfield
  {title} {\emph {\bibinfo {title} {Distributed quantum computation based on
  small quantum registers}},\ }\href
  {https://doi.org/10.1103/PhysRevA.76.062323} {\bibfield  {journal} {\bibinfo
  {journal} {Phys. Rev. A}\ }\textbf {\bibinfo {volume} {76}},\ \bibinfo
  {pages} {062323} (\bibinfo {year} {2007})}\BibitemShut {NoStop}%
\bibitem [{\citenamefont {Qin}\ \emph {et~al.}(2017)\citenamefont {Qin},
  \citenamefont {Wang}, \citenamefont {Miranowicz}, \citenamefont {Zhong},\
  and\ \citenamefont {Nori}}]{qin2017heralded}%
  \BibitemOpen
  \bibfield  {author} {\bibinfo {author} {\bibfnamefont {W.}~\bibnamefont
  {Qin}}, \bibinfo {author} {\bibfnamefont {X.}~\bibnamefont {Wang}}, \bibinfo
  {author} {\bibfnamefont {A.}~\bibnamefont {Miranowicz}}, \bibinfo {author}
  {\bibfnamefont {Z.}~\bibnamefont {Zhong}},\ and\ \bibinfo {author}
  {\bibfnamefont {F.}~\bibnamefont {Nori}},\ }\bibfield  {title} {\emph
  {\bibinfo {title} {Heralded quantum controlled-phase gates with dissipative
  dynamics in macroscopically distant resonators}},\ }\href
  {https://doi.org/https://doi.org/10.1103/PhysRevA.96.012315} {\bibfield
  {journal} {\bibinfo  {journal} {Phys. Rev. A}\ }\textbf {\bibinfo {volume}
  {96}},\ \bibinfo {pages} {012315} (\bibinfo {year} {2017})}\BibitemShut
  {NoStop}%
\bibitem [{\citenamefont {Su}\ \emph {et~al.}(2024)\citenamefont {Su},
  \citenamefont {Qin}, \citenamefont {Miranowicz}, \citenamefont {Li},\ and\
  \citenamefont {Nori}}]{Su2024Heralded}%
  \BibitemOpen
  \bibfield  {author} {\bibinfo {author} {\bibfnamefont {W.}~\bibnamefont
  {Su}}, \bibinfo {author} {\bibfnamefont {W.}~\bibnamefont {Qin}}, \bibinfo
  {author} {\bibfnamefont {A.}~\bibnamefont {Miranowicz}}, \bibinfo {author}
  {\bibfnamefont {T.}~\bibnamefont {Li}},\ and\ \bibinfo {author}
  {\bibfnamefont {F.}~\bibnamefont {Nori}},\ }\bibfield  {title} {\emph
  {\bibinfo {title} {Heralded nonlocal quantum gates for distributed quantum
  computation in a decoherence-free subspace}},\ }\href
  {https://doi.org/10.1103/PhysRevA.110.052612} {\bibfield  {journal} {\bibinfo
   {journal} {Phys. Rev. A}\ }\textbf {\bibinfo {volume} {110}},\ \bibinfo
  {pages} {052612} (\bibinfo {year} {2024})}\BibitemShut {NoStop}%
\bibitem [{\citenamefont {Ruf}\ \emph {et~al.}(2021)\citenamefont {Ruf},
  \citenamefont {Wan}, \citenamefont {Choi}, \citenamefont {Englund},\ and\
  \citenamefont {Hanson}}]{Ruf2021Quantum}%
  \BibitemOpen
  \bibfield  {author} {\bibinfo {author} {\bibfnamefont {M.}~\bibnamefont
  {Ruf}}, \bibinfo {author} {\bibfnamefont {N.~H.}\ \bibnamefont {Wan}},
  \bibinfo {author} {\bibfnamefont {H.}~\bibnamefont {Choi}}, \bibinfo {author}
  {\bibfnamefont {D.}~\bibnamefont {Englund}},\ and\ \bibinfo {author}
  {\bibfnamefont {R.}~\bibnamefont {Hanson}},\ }\bibfield  {title} {\emph
  {\bibinfo {title} {Quantum networks based on color centers in diamond}},\
  }\href {https://doi.org/10.1063/5.0056534} {\bibfield  {journal} {\bibinfo
  {journal} {J. Appl. Phys.}\ }\textbf {\bibinfo {volume} {130}},\ \bibinfo
  {pages} {070901} (\bibinfo {year} {2021})}\BibitemShut {NoStop}%
\bibitem [{\citenamefont {Wehner}\ \emph {et~al.}(2018)\citenamefont {Wehner},
  \citenamefont {Elkouss},\ and\ \citenamefont {Hanson}}]{wehner2018quantum}%
  \BibitemOpen
  \bibfield  {author} {\bibinfo {author} {\bibfnamefont {S.}~\bibnamefont
  {Wehner}}, \bibinfo {author} {\bibfnamefont {D.}~\bibnamefont {Elkouss}},\
  and\ \bibinfo {author} {\bibfnamefont {R.}~\bibnamefont {Hanson}},\
  }\bibfield  {title} {\emph {\bibinfo {title} {Quantum internet: A vision for
  the road ahead}},\ }\href
  {https://science.sciencemag.org/content/362/6412/eaam9288} {\bibfield
  {journal} {\bibinfo  {journal} {Science}\ }\textbf {\bibinfo {volume}
  {362}},\ \bibinfo {pages} {{eaam9288}} (\bibinfo {year} {2018})}\BibitemShut
  {NoStop}%
\bibitem [{\citenamefont {Sheng}\ \emph {et~al.}(2013)\citenamefont {Sheng},
  \citenamefont {Zhou},\ and\ \citenamefont {Long}}]{sheng2013hybrid}%
  \BibitemOpen
  \bibfield  {author} {\bibinfo {author} {\bibfnamefont {Y.-B.}\ \bibnamefont
  {Sheng}}, \bibinfo {author} {\bibfnamefont {L.}~\bibnamefont {Zhou}},\ and\
  \bibinfo {author} {\bibfnamefont {G.-L.}\ \bibnamefont {Long}},\ }\bibfield
  {title} {\emph {\bibinfo {title} {Hybrid entanglement purification for
  quantum repeaters}},\ }\href {https://doi.org/10.1103/PhysRevA.88.022302}
  {\bibfield  {journal} {\bibinfo  {journal} {Phys. Rev. A}\ }\textbf {\bibinfo
  {volume} {88}},\ \bibinfo {pages} {022302} (\bibinfo {year}
  {2013})}\BibitemShut {NoStop}%
\bibitem [{\citenamefont {Qin}\ \emph {et~al.}(2024)\citenamefont {Qin},
  \citenamefont {Kockum}, \citenamefont {{n}oz}, \citenamefont {Miranowicz},\
  and\ \citenamefont {Nori}}]{qin2024quantum}%
  \BibitemOpen
  \bibfield  {author} {\bibinfo {author} {\bibfnamefont {W.}~\bibnamefont
  {Qin}}, \bibinfo {author} {\bibfnamefont {A.~F.}\ \bibnamefont {Kockum}},
  \bibinfo {author} {\bibfnamefont {C.~S.~M.}\ \bibnamefont {{n}oz}}, \bibinfo
  {author} {\bibfnamefont {A.}~\bibnamefont {Miranowicz}},\ and\ \bibinfo
  {author} {\bibfnamefont {F.}~\bibnamefont {Nori}},\ }\bibfield  {title}
  {\emph {\bibinfo {title} {Quantum amplification and simulation of strong and
  ultrastrong coupling of light and matter}},\ }\href
  {https://doi.org/https://doi.org/10.1016/j.physrep.2024.05.003} {\bibfield
  {journal} {\bibinfo  {journal} {Phys. Rep.}\ }\textbf {\bibinfo {volume}
  {1078}},\ \bibinfo {pages} {1} (\bibinfo {year} {2024})}\BibitemShut
  {NoStop}%
\bibitem [{\citenamefont {Xiang}\ \emph {et~al.}(2013)\citenamefont {Xiang},
  \citenamefont {Ashhab}, \citenamefont {You},\ and\ \citenamefont
  {Nori}}]{xiang2013Hybrid}%
  \BibitemOpen
  \bibfield  {author} {\bibinfo {author} {\bibfnamefont {Z.-L.}\ \bibnamefont
  {Xiang}}, \bibinfo {author} {\bibfnamefont {S.}~\bibnamefont {Ashhab}},
  \bibinfo {author} {\bibfnamefont {J.~Q.}\ \bibnamefont {You}},\ and\ \bibinfo
  {author} {\bibfnamefont {F.}~\bibnamefont {Nori}},\ }\bibfield  {title}
  {\emph {\bibinfo {title} {Hybrid quantum circuits: Superconducting circuits
  interacting with other quantum systems}},\ }\href
  {https://doi.org/10.1103/RevModPhys.85.623} {\bibfield  {journal} {\bibinfo
  {journal} {Rev. Mod. Phys.}\ }\textbf {\bibinfo {volume} {85}},\ \bibinfo
  {pages} {623} (\bibinfo {year} {2013})}\BibitemShut {NoStop}%
\bibitem [{\citenamefont {Lodahl}\ \emph {et~al.}(2015)\citenamefont {Lodahl},
  \citenamefont {Mahmoodian},\ and\ \citenamefont
  {Stobbe}}]{lodahl2015interfacing}%
  \BibitemOpen
  \bibfield  {author} {\bibinfo {author} {\bibfnamefont {P.}~\bibnamefont
  {Lodahl}}, \bibinfo {author} {\bibfnamefont {S.}~\bibnamefont {Mahmoodian}},\
  and\ \bibinfo {author} {\bibfnamefont {S.}~\bibnamefont {Stobbe}},\
  }\bibfield  {title} {\emph {\bibinfo {title} {Interfacing single photons and
  single quantum dots with photonic nanostructures}},\ }\href
  {https://link.aps.org/doi/10.1103/RevModPhys.87.347} {\bibfield  {journal}
  {\bibinfo  {journal} {Rev. Mod. Phys.}\ }\textbf {\bibinfo {volume} {87}},\
  \bibinfo {pages} {347} (\bibinfo {year} {2015})}\BibitemShut {NoStop}%
\bibitem [{\citenamefont {Reiserer}\ and\ \citenamefont
  {Rempe}(2015)}]{reiserer2015cavity}%
  \BibitemOpen
  \bibfield  {author} {\bibinfo {author} {\bibfnamefont {A.}~\bibnamefont
  {Reiserer}}\ and\ \bibinfo {author} {\bibfnamefont {G.}~\bibnamefont
  {Rempe}},\ }\bibfield  {title} {\emph {\bibinfo {title} {Cavity-based quantum
  networks with single atoms and optical photons}},\ }\href
  {https://journals.aps.org/rmp/abstract/10.1103/RevModPhys.87.1379} {\bibfield
   {journal} {\bibinfo  {journal} {Rev. Mod. Phys.}\ }\textbf {\bibinfo
  {volume} {87}},\ \bibinfo {pages} {1379} (\bibinfo {year}
  {2015})}\BibitemShut {NoStop}%
\bibitem [{\citenamefont {Zhou}\ \emph {et~al.}(2018)\citenamefont {Zhou},
  \citenamefont {Li}, \citenamefont {Li}, \citenamefont {Li},\ and\
  \citenamefont {Li}}]{zhou2018preparing}%
  \BibitemOpen
  \bibfield  {author} {\bibinfo {author} {\bibfnamefont {Y.}~\bibnamefont
  {Zhou}}, \bibinfo {author} {\bibfnamefont {B.}~\bibnamefont {Li}}, \bibinfo
  {author} {\bibfnamefont {X.-X.}\ \bibnamefont {Li}}, \bibinfo {author}
  {\bibfnamefont {F.-L.}\ \bibnamefont {Li}},\ and\ \bibinfo {author}
  {\bibfnamefont {P.-B.}\ \bibnamefont {Li}},\ }\bibfield  {title} {\emph
  {\bibinfo {title} {Preparing multiparticle entangled states of
  nitrogen-vacancy centers via adiabatic ground-state transitions}},\ }\href
  {https://doi.org/10.1103/PhysRevA.98.052346} {\bibfield  {journal} {\bibinfo
  {journal} {Phys. Rev. A}\ }\textbf {\bibinfo {volume} {98}},\ \bibinfo
  {pages} {052346} (\bibinfo {year} {2018})}\BibitemShut {NoStop}%
\bibitem [{\citenamefont {Qiao}\ \emph {et~al.}(2022)\citenamefont {Qiao},
  \citenamefont {Chen}, \citenamefont {Dong}, \citenamefont {Wang},
  \citenamefont {Hei}, \citenamefont {Shen}, \citenamefont {Zhou},\ and\
  \citenamefont {Li}}]{Qiao2022Generation}%
  \BibitemOpen
  \bibfield  {author} {\bibinfo {author} {\bibfnamefont {Y.-F.}\ \bibnamefont
  {Qiao}}, \bibinfo {author} {\bibfnamefont {J.-Q.}\ \bibnamefont {Chen}},
  \bibinfo {author} {\bibfnamefont {X.-L.}\ \bibnamefont {Dong}}, \bibinfo
  {author} {\bibfnamefont {B.-L.}\ \bibnamefont {Wang}}, \bibinfo {author}
  {\bibfnamefont {X.-L.}\ \bibnamefont {Hei}}, \bibinfo {author} {\bibfnamefont
  {C.-P.}\ \bibnamefont {Shen}}, \bibinfo {author} {\bibfnamefont
  {Y.}~\bibnamefont {Zhou}},\ and\ \bibinfo {author} {\bibfnamefont {P.-B.}\
  \bibnamefont {Li}},\ }\bibfield  {title} {\emph {\bibinfo {title} {Generation
  of {Greenberger-Horne-Zeilinger} states for silicon-vacancy centers using a
  decoherence-free subspace}},\ }\href
  {https://doi.org/10.1103/PhysRevA.105.032415} {\bibfield  {journal} {\bibinfo
   {journal} {Phys. Rev. A}\ }\textbf {\bibinfo {volume} {105}},\ \bibinfo
  {pages} {032415} (\bibinfo {year} {2022})}\BibitemShut {NoStop}%
\bibitem [{\citenamefont {Xia}\ and\ \citenamefont
  {Twamley}(2016)}]{xia2016generating}%
  \BibitemOpen
  \bibfield  {author} {\bibinfo {author} {\bibfnamefont {K.}~\bibnamefont
  {Xia}}\ and\ \bibinfo {author} {\bibfnamefont {J.}~\bibnamefont {Twamley}},\
  }\bibfield  {title} {\emph {\bibinfo {title} {Generating spin squeezing
  states and {Greenberger-Horne-Zeilinger} entanglement using a hybrid
  phonon-spin ensemble in diamond}},\ }\href
  {https://doi.org/10.1103/PhysRevB.94.205118} {\bibfield  {journal} {\bibinfo
  {journal} {Phys. Rev. B}\ }\textbf {\bibinfo {volume} {94}},\ \bibinfo
  {pages} {205118} (\bibinfo {year} {2016})}\BibitemShut {NoStop}%
\bibitem [{\citenamefont {Macr\`{\i}}\ \emph {et~al.}(2018)\citenamefont
  {Macr\`{\i}}, \citenamefont {Nori},\ and\ \citenamefont
  {Kockum}}]{macri2018simple}%
  \BibitemOpen
  \bibfield  {author} {\bibinfo {author} {\bibfnamefont {V.}~\bibnamefont
  {Macr\`{\i}}}, \bibinfo {author} {\bibfnamefont {F.}~\bibnamefont {Nori}},\
  and\ \bibinfo {author} {\bibfnamefont {A.~F.}\ \bibnamefont {Kockum}},\
  }\bibfield  {title} {\emph {\bibinfo {title} {Simple preparation of {Bell and
  Greenberger-Horne-Zeilinger} states using ultrastrong-coupling circuit
  {QED}}},\ }\href {https://doi.org/10.1103/PhysRevA.98.062327} {\bibfield
  {journal} {\bibinfo  {journal} {Phys. Rev. A}\ }\textbf {\bibinfo {volume}
  {98}},\ \bibinfo {pages} {062327} (\bibinfo {year} {2018})}\BibitemShut
  {NoStop}%
\bibitem [{\citenamefont {Northup}\ and\ \citenamefont
  {Blatt}(2014)}]{northup2014quantum}%
  \BibitemOpen
  \bibfield  {author} {\bibinfo {author} {\bibfnamefont {T.}~\bibnamefont
  {Northup}}\ and\ \bibinfo {author} {\bibfnamefont {R.}~\bibnamefont
  {Blatt}},\ }\bibfield  {title} {\emph {\bibinfo {title} {Quantum information
  transfer using photons}},\ }\href
  {https://www.nature.com/articles/nphoton.2014.53} {\bibfield  {journal}
  {\bibinfo  {journal} {Nat. Photonics}\ }\textbf {\bibinfo {volume} {8}},\
  \bibinfo {pages} {356} (\bibinfo {year} {2014})}\BibitemShut {NoStop}%
\bibitem [{\citenamefont {Munro}\ \emph {et~al.}(2015)\citenamefont {Munro},
  \citenamefont {Azuma}, \citenamefont {Tamaki},\ and\ \citenamefont
  {Nemoto}}]{Munro2015Inside}%
  \BibitemOpen
  \bibfield  {author} {\bibinfo {author} {\bibfnamefont {W.~J.}\ \bibnamefont
  {Munro}}, \bibinfo {author} {\bibfnamefont {K.}~\bibnamefont {Azuma}},
  \bibinfo {author} {\bibfnamefont {K.}~\bibnamefont {Tamaki}},\ and\ \bibinfo
  {author} {\bibfnamefont {K.}~\bibnamefont {Nemoto}},\ }\bibfield  {title}
  {\emph {\bibinfo {title} {Inside quantum repeaters}},\ }\href
  {https://doi.org/10.1109/JSTQE.2015.2392076} {\bibfield  {journal} {\bibinfo
  {journal} {IEEE J. Sel. Top. Quantum Electron.}\ }\textbf {\bibinfo {volume}
  {21}},\ \bibinfo {pages} {78} (\bibinfo {year} {2015})}\BibitemShut {NoStop}%
\bibitem [{\citenamefont {Borregaard}\ \emph {et~al.}(2019)\citenamefont
  {Borregaard}, \citenamefont {S{\o}rensen},\ and\ \citenamefont
  {Lodahl}}]{borregaard2019quantum}%
  \BibitemOpen
  \bibfield  {author} {\bibinfo {author} {\bibfnamefont {J.}~\bibnamefont
  {Borregaard}}, \bibinfo {author} {\bibfnamefont {A.~S.}\ \bibnamefont
  {S{\o}rensen}},\ and\ \bibinfo {author} {\bibfnamefont {P.}~\bibnamefont
  {Lodahl}},\ }\bibfield  {title} {\emph {\bibinfo {title} {Quantum networks
  with deterministic spin-photon interfaces}},\ }\href
  {https://doi.org/https://doi.org/10.1002/qute.201800091} {\bibfield
  {journal} {\bibinfo  {journal} {Adv. Quantum Technol.}\ }\textbf {\bibinfo
  {volume} {2}},\ \bibinfo {pages} {1800091} (\bibinfo {year}
  {2019})}\BibitemShut {NoStop}%
\bibitem [{\citenamefont {Cabrillo}\ \emph {et~al.}(1999)\citenamefont
  {Cabrillo}, \citenamefont {Cirac}, \citenamefont {Garc\'{\i}a-Fern\'andez},\
  and\ \citenamefont {Zoller}}]{cabrillo1999creation}%
  \BibitemOpen
  \bibfield  {author} {\bibinfo {author} {\bibfnamefont {C.}~\bibnamefont
  {Cabrillo}}, \bibinfo {author} {\bibfnamefont {J.~I.}\ \bibnamefont {Cirac}},
  \bibinfo {author} {\bibfnamefont {P.}~\bibnamefont
  {Garc\'{\i}a-Fern\'andez}},\ and\ \bibinfo {author} {\bibfnamefont
  {P.}~\bibnamefont {Zoller}},\ }\bibfield  {title} {\emph {\bibinfo {title}
  {Creation of entangled states of distant atoms by interference}},\ }\href
  {https://doi.org/10.1103/PhysRevA.59.1025} {\bibfield  {journal} {\bibinfo
  {journal} {Phys. Rev. A}\ }\textbf {\bibinfo {volume} {59}},\ \bibinfo
  {pages} {1025} (\bibinfo {year} {1999})}\BibitemShut {NoStop}%
\bibitem [{\citenamefont {Yu}\ \emph {et~al.}(2007)\citenamefont {Yu},
  \citenamefont {Yi}, \citenamefont {Song},\ and\ \citenamefont
  {Mei}}]{yu2007robust}%
  \BibitemOpen
  \bibfield  {author} {\bibinfo {author} {\bibfnamefont {C.-S.}\ \bibnamefont
  {Yu}}, \bibinfo {author} {\bibfnamefont {X.~X.}\ \bibnamefont {Yi}}, \bibinfo
  {author} {\bibfnamefont {H.-S.}\ \bibnamefont {Song}},\ and\ \bibinfo
  {author} {\bibfnamefont {D.}~\bibnamefont {Mei}},\ }\bibfield  {title} {\emph
  {\bibinfo {title} {Robust preparation of {Greenberger-Horne-Zeilinger and $W$
  states} of three distant atoms}},\ }\href
  {https://doi.org/10.1103/PhysRevA.75.044301} {\bibfield  {journal} {\bibinfo
  {journal} {Phys. Rev. A}\ }\textbf {\bibinfo {volume} {75}},\ \bibinfo
  {pages} {044301} (\bibinfo {year} {2007})}\BibitemShut {NoStop}%
\bibitem [{\citenamefont {Li}\ and\ \citenamefont
  {Deng}(2016)}]{li2016rejecting}%
  \BibitemOpen
  \bibfield  {author} {\bibinfo {author} {\bibfnamefont {T.}~\bibnamefont
  {Li}}\ and\ \bibinfo {author} {\bibfnamefont {F.-G.}\ \bibnamefont {Deng}},\
  }\bibfield  {title} {\emph {\bibinfo {title} {Error-rejecting quantum
  computing with solid-state spins assisted by {low-\emph{Q}} optical
  microcavities}},\ }\href {https://doi.org/10.1103/PhysRevA.94.062310}
  {\bibfield  {journal} {\bibinfo  {journal} {Phys. Rev. A}\ }\textbf {\bibinfo
  {volume} {94}},\ \bibinfo {pages} {062310} (\bibinfo {year}
  {2016})}\BibitemShut {NoStop}%
\bibitem [{\citenamefont {Nemoto}\ \emph {et~al.}(2014)\citenamefont {Nemoto},
  \citenamefont {Trupke}, \citenamefont {Devitt}, \citenamefont {Stephens},
  \citenamefont {Scharfenberger}, \citenamefont {Buczak}, \citenamefont
  {N\"obauer}, \citenamefont {Everitt}, \citenamefont {Schmiedmayer},\ and\
  \citenamefont {Munro}}]{Nemoto2014Photonic}%
  \BibitemOpen
  \bibfield  {author} {\bibinfo {author} {\bibfnamefont {K.}~\bibnamefont
  {Nemoto}}, \bibinfo {author} {\bibfnamefont {M.}~\bibnamefont {Trupke}},
  \bibinfo {author} {\bibfnamefont {S.~J.}\ \bibnamefont {Devitt}}, \bibinfo
  {author} {\bibfnamefont {A.~M.}\ \bibnamefont {Stephens}}, \bibinfo {author}
  {\bibfnamefont {B.}~\bibnamefont {Scharfenberger}}, \bibinfo {author}
  {\bibfnamefont {K.}~\bibnamefont {Buczak}}, \bibinfo {author} {\bibfnamefont
  {T.}~\bibnamefont {N\"obauer}}, \bibinfo {author} {\bibfnamefont {M.~S.}\
  \bibnamefont {Everitt}}, \bibinfo {author} {\bibfnamefont {J.}~\bibnamefont
  {Schmiedmayer}},\ and\ \bibinfo {author} {\bibfnamefont {W.~J.}\ \bibnamefont
  {Munro}},\ }\bibfield  {title} {\emph {\bibinfo {title} {Photonic
  architecture for scalable quantum information processing in diamond}},\
  }\href {https://doi.org/10.1103/PhysRevX.4.031022} {\bibfield  {journal}
  {\bibinfo  {journal} {Phys. Rev. X}\ }\textbf {\bibinfo {volume} {4}},\
  \bibinfo {pages} {031022} (\bibinfo {year} {2014})}\BibitemShut {NoStop}%
\bibitem [{\citenamefont {Hurst}\ \emph {et~al.}(2019)\citenamefont {Hurst},
  \citenamefont {Joanesarson}, \citenamefont {Iles-Smith}, \citenamefont
  {M\o{}rk},\ and\ \citenamefont {Kok}}]{hurst2019generating}%
  \BibitemOpen
  \bibfield  {author} {\bibinfo {author} {\bibfnamefont {D.~L.}\ \bibnamefont
  {Hurst}}, \bibinfo {author} {\bibfnamefont {K.~B.}\ \bibnamefont
  {Joanesarson}}, \bibinfo {author} {\bibfnamefont {J.}~\bibnamefont
  {Iles-Smith}}, \bibinfo {author} {\bibfnamefont {J.}~\bibnamefont
  {M\o{}rk}},\ and\ \bibinfo {author} {\bibfnamefont {P.}~\bibnamefont {Kok}},\
  }\bibfield  {title} {\emph {\bibinfo {title} {Generating maximal entanglement
  between spectrally distinct solid-state emitters}},\ }\href
  {https://doi.org/10.1103/PhysRevLett.123.023603} {\bibfield  {journal}
  {\bibinfo  {journal} {Phys. Rev. Lett.}\ }\textbf {\bibinfo {volume} {123}},\
  \bibinfo {pages} {023603} (\bibinfo {year} {2019})}\BibitemShut {NoStop}%
\bibitem [{\citenamefont {Pompili}\ \emph {et~al.}(2021)\citenamefont
  {Pompili}, \citenamefont {Hermans}, \citenamefont {Baier}, \citenamefont
  {Beukers}, \citenamefont {Humphreys}, \citenamefont {Schouten}, \citenamefont
  {Vermeulen}, \citenamefont {Tiggelman}, \citenamefont {dos Santos~Martins},
  \citenamefont {Dirkse}, \citenamefont {Wehner},\ and\ \citenamefont
  {Hanson}}]{Pompili2021Realization}%
  \BibitemOpen
  \bibfield  {author} {\bibinfo {author} {\bibfnamefont {M.}~\bibnamefont
  {Pompili}} \emph {et~al.},\ }\bibfield  {title} {\emph {\bibinfo {title}
  {Realization of a multinode quantum network of remote solid-state qubits}},\
  }\href {https://doi.org/10.1126/science.abg1919} {\bibfield  {journal}
  {\bibinfo  {journal} {Science}\ }\textbf {\bibinfo {volume} {372}},\ \bibinfo
  {pages} {259} (\bibinfo {year} {2021})}\BibitemShut {NoStop}%
\bibitem [{\citenamefont {Duan}\ and\ \citenamefont
  {Kimble}(2004)}]{duan2004scalable}%
  \BibitemOpen
  \bibfield  {author} {\bibinfo {author} {\bibfnamefont {L.-M.}\ \bibnamefont
  {Duan}}\ and\ \bibinfo {author} {\bibfnamefont {H.}~\bibnamefont {Kimble}},\
  }\bibfield  {title} {\emph {\bibinfo {title} {Scalable photonic quantum
  computation through cavity-assisted interactions}},\ }\href
  {https://link.aps.org/doi/10.1103/PhysRevLett.92.127902} {\bibfield
  {journal} {\bibinfo  {journal} {Phys. Rev. Lett.}\ }\textbf {\bibinfo
  {volume} {92}},\ \bibinfo {pages} {127902} (\bibinfo {year}
  {2004})}\BibitemShut {NoStop}%
\bibitem [{\citenamefont {Hu}\ \emph {et~al.}(2008)\citenamefont {Hu},
  \citenamefont {Young}, \citenamefont {O'Brien}, \citenamefont {Munro},\ and\
  \citenamefont {Rarity}}]{Hu2008Giant}%
  \BibitemOpen
  \bibfield  {author} {\bibinfo {author} {\bibfnamefont {C.~Y.}\ \bibnamefont
  {Hu}}, \bibinfo {author} {\bibfnamefont {A.}~\bibnamefont {Young}}, \bibinfo
  {author} {\bibfnamefont {J.~L.}\ \bibnamefont {O'Brien}}, \bibinfo {author}
  {\bibfnamefont {W.~J.}\ \bibnamefont {Munro}},\ and\ \bibinfo {author}
  {\bibfnamefont {J.~G.}\ \bibnamefont {Rarity}},\ }\bibfield  {title} {\emph
  {\bibinfo {title} {Giant optical {Faraday} rotation induced by a
  single-electron spin in a quantum dot: Applications to entangling remote
  spins via a single photon}},\ }\href
  {https://doi.org/10.1103/PhysRevB.78.085307} {\bibfield  {journal} {\bibinfo
  {journal} {Phys. Rev. B}\ }\textbf {\bibinfo {volume} {78}},\ \bibinfo
  {pages} {085307} (\bibinfo {year} {2008})}\BibitemShut {NoStop}%
\bibitem [{\citenamefont {Wang}\ \emph {et~al.}(2011)\citenamefont {Wang},
  \citenamefont {Zhang},\ and\ \citenamefont {Jin}}]{WangC2011EP}%
  \BibitemOpen
  \bibfield  {author} {\bibinfo {author} {\bibfnamefont {C.}~\bibnamefont
  {Wang}}, \bibinfo {author} {\bibfnamefont {Y.}~\bibnamefont {Zhang}},\ and\
  \bibinfo {author} {\bibfnamefont {G.-S.}\ \bibnamefont {Jin}},\ }\bibfield
  {title} {\emph {\bibinfo {title} {Entanglement purification and concentration
  of electron-spin entangled states using quantum--dot spins in optical
  microcavities}},\ }\href {https://doi.org/10.1103/PhysRevA.84.032307}
  {\bibfield  {journal} {\bibinfo  {journal} {Phys. Rev. A}\ }\textbf {\bibinfo
  {volume} {84}},\ \bibinfo {pages} {032307} (\bibinfo {year}
  {2011})}\BibitemShut {NoStop}%
\bibitem [{\citenamefont {Li}\ \emph {et~al.}(2018)\citenamefont {Li},
  \citenamefont {Miranowicz}, \citenamefont {Hu}, \citenamefont {Xia},\ and\
  \citenamefont {Nori}}]{li2018gate}%
  \BibitemOpen
  \bibfield  {author} {\bibinfo {author} {\bibfnamefont {T.}~\bibnamefont
  {Li}}, \bibinfo {author} {\bibfnamefont {A.}~\bibnamefont {Miranowicz}},
  \bibinfo {author} {\bibfnamefont {X.}~\bibnamefont {Hu}}, \bibinfo {author}
  {\bibfnamefont {K.}~\bibnamefont {Xia}},\ and\ \bibinfo {author}
  {\bibfnamefont {F.}~\bibnamefont {Nori}},\ }\bibfield  {title} {\emph
  {\bibinfo {title} {Quantum memory and gates using a
  $\mathrm{\ensuremath{\Lambda}}$-type quantum emitter coupled to a chiral
  waveguide}},\ }\href {https://doi.org/10.1103/PhysRevA.97.062318} {\bibfield
  {journal} {\bibinfo  {journal} {Phys. Rev. A}\ }\textbf {\bibinfo {volume}
  {97}},\ \bibinfo {pages} {062318} (\bibinfo {year} {2018})}\BibitemShut
  {NoStop}%
\bibitem [{\citenamefont {Du}\ \emph {et~al.}(2025)\citenamefont {Du},
  \citenamefont {Du}, \citenamefont {Bai},\ and\ \citenamefont
  {Tan}}]{Du2025Heralded}%
  \BibitemOpen
  \bibfield  {author} {\bibinfo {author} {\bibfnamefont {F.-F.}\ \bibnamefont
  {Du}}, \bibinfo {author} {\bibfnamefont {X.-S.}\ \bibnamefont {Du}}, \bibinfo
  {author} {\bibfnamefont {Z.-Y.}\ \bibnamefont {Bai}},\ and\ \bibinfo {author}
  {\bibfnamefont {Q.-L.}\ \bibnamefont {Tan}},\ }\bibfield  {title} {\emph
  {\bibinfo {title} {Heralded deterministic {Knill-Laflamme-Milburn} entanglement
  generation for solid-state emitters via waveguide-assisted photon
  scattering}},\ }\href {https://doi.org/10.1103/371q-1s8h} {\bibfield
  {journal} {\bibinfo  {journal} {Phys. Rev. A}\ }\textbf {\bibinfo {volume}
  {112}},\ \bibinfo {pages} {052440} (\bibinfo {year} {2025})}\BibitemShut
  {NoStop}%
\bibitem [{\citenamefont {Liu}\ \emph {et~al.}(2021)\citenamefont {Liu},
  \citenamefont {Hu}, \citenamefont {Li}, \citenamefont {Li}, \citenamefont
  {Li}, \citenamefont {Liang}, \citenamefont {Zhou}, \citenamefont {Li},\ and\
  \citenamefont {Guo}}]{Liu2021Heralded}%
  \BibitemOpen
  \bibfield  {author} {\bibinfo {author} {\bibfnamefont {X.}~\bibnamefont
  {Liu}}, \bibinfo {author} {\bibfnamefont {J.}~\bibnamefont {Hu}}, \bibinfo
  {author} {\bibfnamefont {Z.-F.}\ \bibnamefont {Li}}, \bibinfo {author}
  {\bibfnamefont {X.}~\bibnamefont {Li}}, \bibinfo {author} {\bibfnamefont
  {P.-Y.}\ \bibnamefont {Li}}, \bibinfo {author} {\bibfnamefont {P.-J.}\
  \bibnamefont {Liang}}, \bibinfo {author} {\bibfnamefont {Z.-Q.}\ \bibnamefont
  {Zhou}}, \bibinfo {author} {\bibfnamefont {C.-F.}\ \bibnamefont {Li}},\ and\
  \bibinfo {author} {\bibfnamefont {G.-C.}\ \bibnamefont {Guo}},\ }\bibfield
  {title} {\emph {\bibinfo {title} {Heralded entanglement distribution between
  two absorptive quantum memories}},\ }\href
  {https://doi.org/10.1038/s41586-021-03505-3} {\bibfield  {journal} {\bibinfo
  {journal} {Nature}\ }\textbf {\bibinfo {volume} {594}},\ \bibinfo {pages}
  {41} (\bibinfo {year} {2021})}\BibitemShut {NoStop}%
\bibitem [{\citenamefont {Tiurev}\ \emph {et~al.}(2021)\citenamefont {Tiurev},
  \citenamefont {Mirambell}, \citenamefont {Lauritzen}, \citenamefont {Appel},
  \citenamefont {Tiranov}, \citenamefont {Lodahl},\ and\ \citenamefont
  {S\o{}rensen}}]{tiurev2021fidelity}%
  \BibitemOpen
  \bibfield  {author} {\bibinfo {author} {\bibfnamefont {K.}~\bibnamefont
  {Tiurev}}, \bibinfo {author} {\bibfnamefont {P.~L.}\ \bibnamefont
  {Mirambell}}, \bibinfo {author} {\bibfnamefont {M.~B.}\ \bibnamefont
  {Lauritzen}}, \bibinfo {author} {\bibfnamefont {M.~H.}\ \bibnamefont
  {Appel}}, \bibinfo {author} {\bibfnamefont {A.}~\bibnamefont {Tiranov}},
  \bibinfo {author} {\bibfnamefont {P.}~\bibnamefont {Lodahl}},\ and\ \bibinfo
  {author} {\bibfnamefont {A.~S.}\ \bibnamefont {S\o{}rensen}},\ }\bibfield
  {title} {\emph {\bibinfo {title} {Fidelity of time-bin-entangled multiphoton
  states from a quantum emitter}},\ }\href
  {https://journals.aps.org/pra/abstract/10.1103/PhysRevA.104.052604}
  {\bibfield  {journal} {\bibinfo  {journal} {Phys. Rev. A}\ }\textbf {\bibinfo
  {volume} {104}},\ \bibinfo {pages} {052604} (\bibinfo {year}
  {2021})}\BibitemShut {NoStop}%
\bibitem [{\citenamefont {Lago-Rivera}\ \emph {et~al.}(2021)\citenamefont
  {Lago-Rivera}, \citenamefont {Grandi}, \citenamefont {Rakonjac},
  \citenamefont {Seri},\ and\ \citenamefont
  {de~Riedmatten}}]{Lago-Rivera2021Telecom-heralded}%
  \BibitemOpen
  \bibfield  {author} {\bibinfo {author} {\bibfnamefont {D.}~\bibnamefont
  {Lago-Rivera}}, \bibinfo {author} {\bibfnamefont {S.}~\bibnamefont {Grandi}},
  \bibinfo {author} {\bibfnamefont {J.~V.}\ \bibnamefont {Rakonjac}}, \bibinfo
  {author} {\bibfnamefont {A.}~\bibnamefont {Seri}},\ and\ \bibinfo {author}
  {\bibfnamefont {H.}~\bibnamefont {de~Riedmatten}},\ }\bibfield  {title}
  {\emph {\bibinfo {title} {Telecom-heralded entanglement between multimode
  solid-state quantum memories}},\ }\href
  {https://doi.org/10.1038/s41586-021-03481-8} {\bibfield  {journal} {\bibinfo
  {journal} {Nature}\ }\textbf {\bibinfo {volume} {594}},\ \bibinfo {pages}
  {37} (\bibinfo {year} {2021})}\BibitemShut {NoStop}%
\bibitem [{\citenamefont {Jones}\ \emph {et~al.}(2016)\citenamefont {Jones},
  \citenamefont {Kim}, \citenamefont {Rakher}, \citenamefont {Kwiat},\ and\
  \citenamefont {Ladd}}]{Jones2016Design}%
  \BibitemOpen
  \bibfield  {author} {\bibinfo {author} {\bibfnamefont {C.}~\bibnamefont
  {Jones}}, \bibinfo {author} {\bibfnamefont {D.}~\bibnamefont {Kim}}, \bibinfo
  {author} {\bibfnamefont {M.~T.}\ \bibnamefont {Rakher}}, \bibinfo {author}
  {\bibfnamefont {P.~G.}\ \bibnamefont {Kwiat}},\ and\ \bibinfo {author}
  {\bibfnamefont {T.~D.}\ \bibnamefont {Ladd}},\ }\bibfield  {title} {\emph
  {\bibinfo {title} {Design and analysis of communication protocols for quantum
  repeater networks}},\ }\href {https://doi.org/10.1088/1367-2630/18/8/083015}
  {\bibfield  {journal} {\bibinfo  {journal} {New J. Phys.}\ }\textbf {\bibinfo
  {volume} {18}},\ \bibinfo {pages} {083015} (\bibinfo {year}
  {2016})}\BibitemShut {NoStop}%
\bibitem [{\citenamefont {Liu}\ and\ \citenamefont
  {Wei}(2025)}]{Liu2025Deterministic}%
  \BibitemOpen
  \bibfield  {author} {\bibinfo {author} {\bibfnamefont {W.-Q.}\ \bibnamefont
  {Liu}}\ and\ \bibinfo {author} {\bibfnamefont {H.-R.}\ \bibnamefont {Wei}},\
  }\bibfield  {title} {\emph {\bibinfo {title} {Deterministic generation of
  multiqubit entangled states among distant parties using indefinite causal
  order}},\ }\href {https://doi.org/10.1103/PhysRevApplied.23.054075}
  {\bibfield  {journal} {\bibinfo  {journal} {Phys. Rev. Appl.}\ }\textbf
  {\bibinfo {volume} {23}},\ \bibinfo {pages} {054075} (\bibinfo {year}
  {2025})}\BibitemShut {NoStop}%
\bibitem [{\citenamefont {Lo~Piparo}\ \emph {et~al.}(2019)\citenamefont
  {Lo~Piparo}, \citenamefont {Munro},\ and\ \citenamefont
  {Nemoto}}]{Piparo2019multiplexing}%
  \BibitemOpen
  \bibfield  {author} {\bibinfo {author} {\bibfnamefont {N.}~\bibnamefont
  {Lo~Piparo}}, \bibinfo {author} {\bibfnamefont {W.~J.}\ \bibnamefont
  {Munro}},\ and\ \bibinfo {author} {\bibfnamefont {K.}~\bibnamefont
  {Nemoto}},\ }\bibfield  {title} {\emph {\bibinfo {title} {Quantum
  multiplexing}},\ }\href {https://doi.org/10.1103/PhysRevA.99.022337}
  {\bibfield  {journal} {\bibinfo  {journal} {Phys. Rev. A}\ }\textbf {\bibinfo
  {volume} {99}},\ \bibinfo {pages} {022337} (\bibinfo {year}
  {2019})}\BibitemShut {NoStop}%
\bibitem [{\citenamefont {Xie}\ \emph {et~al.}(2021)\citenamefont {Xie},
  \citenamefont {Liu}, \citenamefont {Mo}, \citenamefont {Li},\ and\
  \citenamefont {Li}}]{xie2021quantum}%
  \BibitemOpen
  \bibfield  {author} {\bibinfo {author} {\bibfnamefont {Z.}~\bibnamefont
  {Xie}}, \bibinfo {author} {\bibfnamefont {Y.}~\bibnamefont {Liu}}, \bibinfo
  {author} {\bibfnamefont {X.}~\bibnamefont {Mo}}, \bibinfo {author}
  {\bibfnamefont {T.}~\bibnamefont {Li}},\ and\ \bibinfo {author}
  {\bibfnamefont {Z.}~\bibnamefont {Li}},\ }\bibfield  {title} {\emph {\bibinfo
  {title} {Quantum entanglement creation for distant quantum memories via
  time-bin multiplexing}},\ }\href
  {https://doi.org/10.1103/PhysRevA.104.062409} {\bibfield  {journal} {\bibinfo
   {journal} {Phys. Rev. A}\ }\textbf {\bibinfo {volume} {104}},\ \bibinfo
  {pages} {062409} (\bibinfo {year} {2021})}\BibitemShut {NoStop}%
\bibitem [{\citenamefont {Wang}\ \emph {et~al.}(2025)\citenamefont {Wang},
  \citenamefont {Ye}, \citenamefont {Zhang}, \citenamefont {Wei},\ and\
  \citenamefont {Song}}]{Wang2025Heralded}%
  \BibitemOpen
  \bibfield  {author} {\bibinfo {author} {\bibfnamefont {L.-X.}\ \bibnamefont
  {Wang}}, \bibinfo {author} {\bibfnamefont {Y.-P.}\ \bibnamefont {Ye}},
  \bibinfo {author} {\bibfnamefont {C.-F.}\ \bibnamefont {Zhang}}, \bibinfo
  {author} {\bibfnamefont {H.-R.}\ \bibnamefont {Wei}},\ and\ \bibinfo {author}
  {\bibfnamefont {G.-Z.}\ \bibnamefont {Song}},\ }\bibfield  {title} {\emph
  {\bibinfo {title} {Heralded long-distance entanglement schemes for waveguide
  systems in quantum networks}},\ }\href {https://doi.org/10.1103/ywz1-gqzg}
  {\bibfield  {journal} {\bibinfo  {journal} {Phys. Rev. Appl.}\ }\textbf
  {\bibinfo {volume} {24}},\ \bibinfo {pages} {044070} (\bibinfo {year}
  {2025})}\BibitemShut {NoStop}%
\bibitem [{\citenamefont {Erhard}\ \emph {et~al.}(2020)\citenamefont {Erhard},
  \citenamefont {Krenn},\ and\ \citenamefont
  {Zeilinger}}]{Erhard2020high-dimensional}%
  \BibitemOpen
  \bibfield  {author} {\bibinfo {author} {\bibfnamefont {M.}~\bibnamefont
  {Erhard}}, \bibinfo {author} {\bibfnamefont {M.}~\bibnamefont {Krenn}},\ and\
  \bibinfo {author} {\bibfnamefont {A.}~\bibnamefont {Zeilinger}},\ }\bibfield
  {title} {\emph {\bibinfo {title} {Advances in high-dimensional quantum
  entanglement}},\ }\href {https://doi.org/10.1038/s42254-020-0193-5}
  {\bibfield  {journal} {\bibinfo  {journal} {Nat. Rev. Phys.}\ }\textbf
  {\bibinfo {volume} {2}},\ \bibinfo {pages} {365} (\bibinfo {year}
  {2020})}\BibitemShut {NoStop}%
\bibitem [{\citenamefont {McIntyre}\ and\ \citenamefont
  {Coish}(2025)}]{McIntyre2025Loss-tolerant}%
  \BibitemOpen
  \bibfield  {author} {\bibinfo {author} {\bibfnamefont {Z.~M.}\ \bibnamefont
  {McIntyre}}\ and\ \bibinfo {author} {\bibfnamefont {W.~A.}\ \bibnamefont
  {Coish}},\ }\bibfield  {title} {\emph {\bibinfo {title} {Loss-tolerant
  parallelized {Bell}-state generation with a hybrid cat qudit}},\ }\href
  {https://doi.org/10.1103/x56x-vld7} {\bibfield  {journal} {\bibinfo
  {journal} {Phys. Rev. A}\ }\textbf {\bibinfo {volume} {112}},\ \bibinfo
  {pages} {062609} (\bibinfo {year} {2025})}\BibitemShut {NoStop}%
\bibitem [{\citenamefont {Zheng}\ \emph {et~al.}(2022)\citenamefont {Zheng},
  \citenamefont {Sharma},\ and\ \citenamefont
  {Borregaard}}]{zheng2022entanglement}%
  \BibitemOpen
  \bibfield  {author} {\bibinfo {author} {\bibfnamefont {Y.}~\bibnamefont
  {Zheng}}, \bibinfo {author} {\bibfnamefont {H.}~\bibnamefont {Sharma}},\ and\
  \bibinfo {author} {\bibfnamefont {J.}~\bibnamefont {Borregaard}},\ }\bibfield
   {title} {\emph {\bibinfo {title} {Entanglement distribution with minimal
  memory requirements using time-bin photonic qudits}},\ }\href
  {https://doi.org/10.1103/PRXQuantum.3.040319} {\bibfield  {journal} {\bibinfo
   {journal} {PRX Quantum}\ }\textbf {\bibinfo {volume} {3}},\ \bibinfo {pages}
  {040319} (\bibinfo {year} {2022})}\BibitemShut {NoStop}%
\bibitem [{\citenamefont {Liu}\ \emph {et~al.}(2024)\citenamefont {Liu},
  \citenamefont {Bharos}, \citenamefont {Markovich},\ and\ \citenamefont
  {Borregaard}}]{liu2024error}%
  \BibitemOpen
  \bibfield  {author} {\bibinfo {author} {\bibfnamefont {X.}~\bibnamefont
  {Liu}}, \bibinfo {author} {\bibfnamefont {N.}~\bibnamefont {Bharos}},
  \bibinfo {author} {\bibfnamefont {L.}~\bibnamefont {Markovich}},\ and\
  \bibinfo {author} {\bibfnamefont {J.}~\bibnamefont {Borregaard}},\ }\bibfield
   {title} {\emph {\bibinfo {title} {Error correlations in photonic
  qudit-mediated entanglement generation}},\ }\href
  {https://doi.org/10.1103/PhysRevResearch.6.023075} {\bibfield  {journal}
  {\bibinfo  {journal} {Phys. Rev. Res.}\ }\textbf {\bibinfo {volume} {6}},\
  \bibinfo {pages} {023075} (\bibinfo {year} {2024})}\BibitemShut {NoStop}%
\bibitem [{\citenamefont {Zhou}\ \emph {et~al.}(2023)\citenamefont {Zhou},
  \citenamefont {Li},\ and\ \citenamefont {Xia}}]{zhou2023parallel}%
  \BibitemOpen
  \bibfield  {author} {\bibinfo {author} {\bibfnamefont {H.}~\bibnamefont
  {Zhou}}, \bibinfo {author} {\bibfnamefont {T.}~\bibnamefont {Li}},\ and\
  \bibinfo {author} {\bibfnamefont {K.}~\bibnamefont {Xia}},\ }\bibfield
  {title} {\emph {\bibinfo {title} {Parallel and heralded multiqubit
  entanglement generation for quantum networks}},\ }\href
  {https://doi.org/10.1103/PhysRevA.107.022428} {\bibfield  {journal} {\bibinfo
   {journal} {Phys. Rev. A}\ }\textbf {\bibinfo {volume} {107}},\ \bibinfo
  {pages} {022428} (\bibinfo {year} {2023})}\BibitemShut {NoStop}%
\bibitem [{\citenamefont {Aharonovich}\ \emph {et~al.}(2016)\citenamefont
  {Aharonovich}, \citenamefont {Englund},\ and\ \citenamefont
  {Toth}}]{aharonovich2016solid}%
  \BibitemOpen
  \bibfield  {author} {\bibinfo {author} {\bibfnamefont {I.}~\bibnamefont
  {Aharonovich}}, \bibinfo {author} {\bibfnamefont {D.}~\bibnamefont
  {Englund}},\ and\ \bibinfo {author} {\bibfnamefont {M.}~\bibnamefont
  {Toth}},\ }\bibfield  {title} {\emph {\bibinfo {title} {Solid-state
  single-photon emitters}},\ }\href
  {https://www.nature.com/articles/nphoton.2016.186} {\bibfield  {journal}
  {\bibinfo  {journal} {Nat. Photonics}\ }\textbf {\bibinfo {volume} {10}},\
  \bibinfo {pages} {631} (\bibinfo {year} {2016})}\BibitemShut {NoStop}%
\bibitem [{\citenamefont {Borregaard}\ \emph {et~al.}(2020)\citenamefont
  {Borregaard}, \citenamefont {Pichler}, \citenamefont {Schr\"oder},
  \citenamefont {Lukin}, \citenamefont {Lodahl},\ and\ \citenamefont
  {S\o{}rensen}}]{Borregaard2020One-Way}%
  \BibitemOpen
  \bibfield  {author} {\bibinfo {author} {\bibfnamefont {J.}~\bibnamefont
  {Borregaard}}, \bibinfo {author} {\bibfnamefont {H.}~\bibnamefont {Pichler}},
  \bibinfo {author} {\bibfnamefont {T.}~\bibnamefont {Schr\"oder}}, \bibinfo
  {author} {\bibfnamefont {M.~D.}\ \bibnamefont {Lukin}}, \bibinfo {author}
  {\bibfnamefont {P.}~\bibnamefont {Lodahl}},\ and\ \bibinfo {author}
  {\bibfnamefont {A.~S.}\ \bibnamefont {S\o{}rensen}},\ }\bibfield  {title}
  {\emph {\bibinfo {title} {One-way quantum repeater based on
  near-deterministic photon-emitter interfaces}},\ }\href
  {https://doi.org/10.1103/PhysRevX.10.021071} {\bibfield  {journal} {\bibinfo
  {journal} {Phys. Rev. X}\ }\textbf {\bibinfo {volume} {10}},\ \bibinfo
  {pages} {021071} (\bibinfo {year} {2020})}\BibitemShut {NoStop}%
\bibitem [{\citenamefont {Wang}\ \emph {et~al.}(2018)\citenamefont {Wang},
  \citenamefont {Zhang}, \citenamefont {Chen}, \citenamefont {Bertrand},
  \citenamefont {Shams-Ansari}, \citenamefont {Chandrasekhar}, \citenamefont
  {Winzer},\ and\ \citenamefont {Lon\u{c}ar}}]{Wang2018Integrated}%
  \BibitemOpen
  \bibfield  {author} {\bibinfo {author} {\bibfnamefont {C.}~\bibnamefont
  {Wang}}, \bibinfo {author} {\bibfnamefont {M.}~\bibnamefont {Zhang}},
  \bibinfo {author} {\bibfnamefont {X.}~\bibnamefont {Chen}}, \bibinfo {author}
  {\bibfnamefont {M.}~\bibnamefont {Bertrand}}, \bibinfo {author}
  {\bibfnamefont {A.}~\bibnamefont {Shams-Ansari}}, \bibinfo {author}
  {\bibfnamefont {S.}~\bibnamefont {Chandrasekhar}}, \bibinfo {author}
  {\bibfnamefont {P.}~\bibnamefont {Winzer}},\ and\ \bibinfo {author}
  {\bibfnamefont {M.}~\bibnamefont {Lon\u{c}ar}},\ }\bibfield  {title} {\emph
  {\bibinfo {title} {Integrated lithium niobate electro-optic modulators
  operating at {CMOS}-compatible voltages}},\ }\href
  {https://doi.org/10.1038/s41586-018-0551-y} {\bibfield  {journal} {\bibinfo
  {journal} {{Nature}}\ }\textbf {\bibinfo {volume} {562}},\ \bibinfo {pages}
  {101} (\bibinfo {year} {2018})}\BibitemShut {NoStop}%
\bibitem [{\citenamefont {Knaut}\ \emph {et~al.}(2024)\citenamefont {Knaut},
  \citenamefont {Suleymanzade}, \citenamefont {Wei}, \citenamefont {Assumpcao},
  \citenamefont {Stas}, \citenamefont {Huan}, \citenamefont {Machielse},
  \citenamefont {Knall}, \citenamefont {Sutula}, \citenamefont {Baranes},
  \citenamefont {Sinclair}, \citenamefont {De-Eknamkul}, \citenamefont
  {Levonian}, \citenamefont {Bhaskar}, \citenamefont {Park}, \citenamefont
  {Lončar},\ and\ \citenamefont {Lukin}}]{Knaut2024Entanglement}%
  \BibitemOpen
  \bibfield  {author} {\bibinfo {author} {\bibfnamefont {C.~M.}\ \bibnamefont
  {Knaut}} \emph {et~al.},\ }\bibfield  {title} {\emph {\bibinfo {title}
  {Entanglement of nanophotonic quantum memory nodes in a telecom network}},\
  }\href {https://doi.org/10.1038/s41586-024-07252-z} {\bibfield  {journal}
  {\bibinfo  {journal} {Nature (London)}\ }\textbf {\bibinfo {volume} {629}},\
  \bibinfo {pages} {573} (\bibinfo {year} {2024})}\BibitemShut {NoStop}%
\bibitem [{\citenamefont {Johansson}\ \emph {et~al.}(2013)\citenamefont
  {Johansson}, \citenamefont {Nation},\ and\ \citenamefont
  {Nori}}]{johansson2013qutip}%
  \BibitemOpen
  \bibfield  {author} {\bibinfo {author} {\bibfnamefont {J.}~\bibnamefont
  {Johansson}}, \bibinfo {author} {\bibfnamefont {P.}~\bibnamefont {Nation}},\
  and\ \bibinfo {author} {\bibfnamefont {F.}~\bibnamefont {Nori}},\ }\bibfield
  {title} {\emph {\bibinfo {title} {Qutip 2: A {Python} framework for the
  dynamics of open quantum systems}},\ }\href
  {https://doi.org/https://doi.org/10.1016/j.cpc.2012.11.019} {\bibfield
  {journal} {\bibinfo  {journal} {Comput. Phys. Commun.}\ }\textbf {\bibinfo
  {volume} {184}},\ \bibinfo {pages} {1234} (\bibinfo {year}
  {2013})}\BibitemShut {NoStop}%
\bibitem [{\citenamefont {Zhang}\ \emph {et~al.}(2021)\citenamefont {Zhang},
  \citenamefont {Tao}, \citenamefont {He}, \citenamefont {Chen}, \citenamefont
  {Kong}, \citenamefont {Deng}, \citenamefont {Lambert},\ and\ \citenamefont
  {Ai}}]{Zhang2021Efficient}%
  \BibitemOpen
  \bibfield  {author} {\bibinfo {author} {\bibfnamefont {N.-N.}\ \bibnamefont
  {Zhang}}, \bibinfo {author} {\bibfnamefont {M.-J.}\ \bibnamefont {Tao}},
  \bibinfo {author} {\bibfnamefont {W.-T.}\ \bibnamefont {He}}, \bibinfo
  {author} {\bibfnamefont {X.-Y.}\ \bibnamefont {Chen}}, \bibinfo {author}
  {\bibfnamefont {X.-Y.}\ \bibnamefont {Kong}}, \bibinfo {author}
  {\bibfnamefont {F.-G.}\ \bibnamefont {Deng}}, \bibinfo {author}
  {\bibfnamefont {N.}~\bibnamefont {Lambert}},\ and\ \bibinfo {author}
  {\bibfnamefont {Q.}~\bibnamefont {Ai}},\ }\bibfield  {title} {\emph {\bibinfo
  {title} {Efficient quantum simulation of open quantum dynamics at various
  {Hamiltonians} and spectral densities}},\ }\href
  {https://doi.org/10.1007/s11467-021-1064-y} {\bibfield  {journal} {\bibinfo
  {journal} {Front. Phys.}\ }\textbf {\bibinfo {volume} {16}},\ \bibinfo
  {pages} {51501} (\bibinfo {year} {2021})}\BibitemShut {NoStop}%
\bibitem [{\citenamefont {Guo}\ \emph {et~al.}(2023)\citenamefont {Guo},
  \citenamefont {Xie}, \citenamefont {Wang}, \citenamefont {Li},\ and\
  \citenamefont {Li}}]{Guo2023Heralded}%
  \BibitemOpen
  \bibfield  {author} {\bibinfo {author} {\bibfnamefont {Z.}~\bibnamefont
  {Guo}}, \bibinfo {author} {\bibfnamefont {Z.}~\bibnamefont {Xie}}, \bibinfo
  {author} {\bibfnamefont {Y.}~\bibnamefont {Wang}}, \bibinfo {author}
  {\bibfnamefont {Z.}~\bibnamefont {Li}},\ and\ \bibinfo {author}
  {\bibfnamefont {T.}~\bibnamefont {Li}},\ }\bibfield  {title} {\emph {\bibinfo
  {title} {Heralded and robust {W}-state generation for distant superconducting
  qubits with practical microwave pulse scattering}},\ }\href
  {https://doi.org/10.1063/5.0189377} {\bibfield  {journal} {\bibinfo
  {journal} {Appl. Phys. Lett.}\ }\textbf {\bibinfo {volume} {123}},\ \bibinfo
  {pages} {264002} (\bibinfo {year} {2023})}\BibitemShut {NoStop}%
\bibitem [{\citenamefont {Fitzke}\ \emph {et~al.}(2022)\citenamefont {Fitzke},
  \citenamefont {Bialowons}, \citenamefont {Dolejsky}, \citenamefont
  {Tippmann}, \citenamefont {Nikiforov}, \citenamefont {Walther}, \citenamefont
  {Wissel},\ and\ \citenamefont {Gunkel}}]{Fitzke2022Scalable}%
  \BibitemOpen
  \bibfield  {author} {\bibinfo {author} {\bibfnamefont {E.}~\bibnamefont
  {Fitzke}}, \bibinfo {author} {\bibfnamefont {L.}~\bibnamefont {Bialowons}},
  \bibinfo {author} {\bibfnamefont {T.}~\bibnamefont {Dolejsky}}, \bibinfo
  {author} {\bibfnamefont {M.}~\bibnamefont {Tippmann}}, \bibinfo {author}
  {\bibfnamefont {O.}~\bibnamefont {Nikiforov}}, \bibinfo {author}
  {\bibfnamefont {T.}~\bibnamefont {Walther}}, \bibinfo {author} {\bibfnamefont
  {F.}~\bibnamefont {Wissel}},\ and\ \bibinfo {author} {\bibfnamefont
  {M.}~\bibnamefont {Gunkel}},\ }\bibfield  {title} {\emph {\bibinfo {title}
  {Scalable network for simultaneous pairwise quantum key distribution via
  entanglement-based time-bin coding}},\ }\href
  {https://doi.org/10.1103/PRXQuantum.3.020341} {\bibfield  {journal} {\bibinfo
   {journal} {PRX Quantum}\ }\textbf {\bibinfo {volume} {3}},\ \bibinfo {pages}
  {020341} (\bibinfo {year} {2022})}\BibitemShut {NoStop}%
\bibitem [{\citenamefont {Song}\ \emph {et~al.}(2019)\citenamefont {Song},
  \citenamefont {Xu}, \citenamefont {Li}, \citenamefont {Zhang}, \citenamefont
  {Zhang}, \citenamefont {Liu}, \citenamefont {Guo}, \citenamefont {Wang},
  \citenamefont {Ren}, \citenamefont {Hao}, \citenamefont {Feng}, \citenamefont
  {Fan}, \citenamefont {Zheng}, \citenamefont {Wang}, \citenamefont {Wang},\
  and\ \citenamefont {Zhu}}]{Song2019Generation}%
  \BibitemOpen
  \bibfield  {author} {\bibinfo {author} {\bibfnamefont {C.}~\bibnamefont
  {Song}} \emph {et~al.},\ }\bibfield  {title} {\emph {\bibinfo {title}
  {Generation of multicomponent atomic {Schr\"{o}dinger} cat states of up to 20
  qubits}},\ }\href {https://doi.org/10.1126/science.aay0600} {\bibfield
  {journal} {\bibinfo  {journal} {Science}\ }\textbf {\bibinfo {volume}
  {365}},\ \bibinfo {pages} {574} (\bibinfo {year} {2019})}\BibitemShut
  {NoStop}%
\end{thebibliography}

%

\end{document}